\documentclass[12pt]{iopart}
\usepackage{iopams,psfig,cite}
\eqnobysec

\begin{document}
\jl{1}

\title[Correlation analysis for the CS model]{Hellmann-Feynman
  theorem and correlation-fluctuation analysis for the
  Calogero-Sutherland model}

\author{Rudolf A R\"omer\footnote{{r.roemer@physik.tu-chemnitz.de}}}
\address{Institut f\"ur Physik, Technische Universit\"at, D-09107 Chemnitz, Germany}

\author{Paul Ziesche}
\address{Max-Planck-Institut f\"ur Physik komplexer Systeme,
N\"othnitzer Str. 38, D-01187 Dresden, Germany}

\date{$Revision: 2.26 $; printed \today}

\begin{abstract}
  Exploiting the results of the exact solution for the ground state of
  the one-dimensional spinless quantum gas of Fermions and
  impenetrable Bosons with the $\mu/x_{ij}^2$ particle-particle
  interaction, the Hellmann-Feynman theorem yields mutually
  compensating divergences of both the kinetic and the interaction
  energy in the limiting case $\mu {\to}- {1/4}$.  These divergences
  result from the peculiar behavior of both the momentum distribution
  (for large momenta) and the pair density (for small inter-particle
  separation). The available analytical pair densities for $\mu=-1/4,\
  0,\ {\rm and}\ 2$ allow to analyze particle-number fluctuations.
  They are suppressed by repulsive interaction ($\mu>0$), enhanced by
  attraction ($\mu<0$), and may therefore measure the kind and
  strength of correlation. Other recently proposed purely
  quantum-kinematical measures of the correlation strength arise from
  the small-separation behavior of the pair density or --- for
  Fermions --- from the non-idempotency of the momentum distribution
  and its large-momenta behavior. They are compared with each other
  and with reference-free, short-range correlation-measuring ratios of
  the kinetic and potential energies.


\end{abstract}


\submitted

\maketitle

\section{Introduction}
\label{sec:intro}

In the ground state of electron systems, it has been shown that
exchange (X) due to the Pauli `repulsion' and correlation (C) due to
the Coulomb repulsion suppress particle-number fluctuations and
consequently reduce the energy \cite{Ful95,SchFD98,ZieTSP99}. This
energy reduction provides most of the `glue' that binds atoms together
to form molecules and solids \cite{Ku00}. Particle-number fluctuations
mean that the particle number in a domain (which may be a muffin-tin
sphere, a Wigner-Seitz cell, a Bader basin \cite{Ba90}, a Daudel loge
\cite{As72}, a bond region between atoms in a molecule, etc.)
fluctuates due to zero temperature quantum motion with a certain
probability.  Fulde \cite{Ful95} takes C$_2$H$_2$ as an example for
such fluctuations.  The number of valence electrons in a sphere
containing a C atom fluctuates around its average value $\approx 3.9$.
Comparison of Hartree-Fock (HF) calculations for C$_2$H$_2$ with
calculations which include correlation shows that configurations with
large deviations from the average valence electron number (e.g., with
$0$ and $1$ or $7$ and $8$ electrons) are strongly suppressed due to
correlation. A similar fluctuation-correlation analysis is performed
in Ref.\ \cite{SchFD98} for several dimers and in Ref.\ 
\cite{ZieTSP99} for the uniform electron gas in one, two, and three
dimensions (1D, 2D, 3D).  These calculations for the above mentioned
narrowing of the particle-number distribution need the pair density
(PD) $n({\vec r}_1,{\vec r}_2)$ and this narrowing is used to derive
from the PD a quantum-kinematical measure for the correlation strength
\cite{Ful95}.  Correlation and its strength is furthermore
characterized by the small-separation (or on-top) behavior of the PD.
The spherically averaged on-top curvature of the spin-parallel PD may
serve as a local correlation measure \cite{Dob91} and from the
topological analysis of the intracule PD a short-range correlation
strength is defined \cite{CioL99}. In addition to these PD based
quantities the concept of a correlation `entropy' has been developed
for Fermi systems \cite{Low55,Col93,RamPSE98,ZieGJB97} (in Ref.\ 
\cite{RamPSE98} the term Jaynes entropy is used). It is based on the
correlation induced non-idempotency of the correlated one-particle
density matrix (1PDM) $\gamma({\vec r};{\vec r}')$. All these
correlation measures intend to make the qualitative terms `weak and
strong correlation' quantitatively precise \cite{Zie00a}.  Note that
strong correlation means extreme narrowing of the particle-number
distribution which is usually described as electron localization.

In the present paper, we apply the above mentioned
fluctuation-correlation analysis to the exactly solvable
Calogero-Sutherland (CS) model \cite{Sut71}. The CS model is a model of
long-range-interacting spinless particles in 1D and has been solved
exactly by means of the (asymptotic) Bethe-Ansatz
technique \cite{Sut71,RomS93}. The solution is valid for both fermionic
and bosonic particle symmetry. Here we will mostly concentrate on the
Fermi systems.  Furthermore, the model can be shown to be the
universal quantum model underlying the dynamical interpretation of
random matrix theory \cite{Meh90,Dys62}. This latter connection has
been used to also compute several correlation functions exactly at
three special values of the interaction strength, among them the 1PDM
and the PD \cite{Sut71}. Thus although the information is restricted to
the 1D case, the model nevertheless is ideally suited for testing the
fluctuation-correlation measures discussed above.

Correlated 1PDM and correlated PD need correlated many-body wave
functions (beyond the HF caricature), which in quantum
chemistry \cite{SzaO82,HamLR94} are traditionally obtained from
configuration interaction (CI), coupled cluster (CC),
M{\o}ller-Plesset, quantum Monte Carlo calculations or recently from
the contracted-Schr\"odinger-equation method \cite{Maz99} or the
incremental method \cite{FulSK99}. All these procedures involve certain
approximations or have restricted applicability.  So the existence of
non-trivial {\em exactly solvable models} which can provide 1PDM and
PD is of much interest for the mentioned fluctuation and correlation
analysis.

We shall only consider the ground state properties of the CS
model \cite{Sut71}. The interaction is pairwise inversely proportional
to the square of the distance $x_{ij}=|x_i-x_j|$ of two particles with
interaction strength $\mu$, {\em i.e.}, $\mu/x_{ij}^2$. The
interaction strength $\mu\geq -1/4$ is occasionally parametrized as
$\mu=\lambda (\lambda -1)$ with a parameter $\lambda
=1/2+\sqrt{1/4+\mu} \geq 1/2$. We shall mostly use the
parameterization $\nu=\sqrt{1/4 +\mu}$, such that $\mu = \nu^2-1/4$
and $\lambda =1/2 + \nu$. In the thermodynamic limit we assume
constant density $\rho (x)=n$, so the CS-ground state has only two
parameters, $\nu$ and $n$. The $1/x_{ij}^2$ interaction has the
peculiarity not to possess a natural length. Therefore it is a model
showing critical behavior, which can be discussed in terms of
universality classes and their conformal
anomalies \cite{Pol70,BelPZ84,Car84,Car87}. This beauty of the
$1/x_{ij}^2$ interaction shows up also in the analytical Bethe-Ansatz
solutions \cite{Sut71,SutR93,SutRS94,SimA93a,SimA93b,KraT00} and the
explicit knowledge of the correlated many-body wave
functions \cite{Sut71,OlsP81}.
\mbox{}From the Bethe-Ansatz technique the complete energy spectrum
and in particular the ground state energy per particle as a function
of the interaction strength parameter $\nu$ is available \cite{Sut71}
We show that its kinetic and interaction `components' can be deduced
with the help of the Hellmann-Feynman theorem \cite{Hel37}.
Surprisingly, when the interaction strength parameter $\nu$ approaches
its limiting value $0$, both the kinetic and the interaction energy
diverge in such a way that they compensate each other leaving the
total energy finite.  As we outline in the following, these
divergences result from the peculiar behavior of the 1PDM and the PD
for $\nu\to 0$ and are related to the ``fall-into-the-origin'' already
mentioned in Ref.\ \cite{LanL77}.

For $\nu=0, 1/2$, and $3/2$ --- corresponding to $\mu = - 1/4, 0$ and
$2$ or $\lambda=1/2, 1$, and $2$ --- it has been shown \cite{Sut71}
that the square of the ground state wave function is intimately
related to the eigenvalue distribution of random matrices of the
orthogonal ensemble, the unitary ensemble, and the symplectic
ensemble, respectively.  Using this connection, Sutherland had shown
how to construct the 1PDM and the PD using integral relations of
random matrix theory.  The resulting formulas reduce the problem, say
for the 1PDM, from the evaluation of a high-dimensional integral to
the computation of a determinant of a matrix \cite{RomS93}. From the
1PDM $\gamma (x-x')$, the momentum distributions $n_\kappa $ for the
three special values of $\nu$ follow via Fourier transform.  Due to
correlation the latter quantities are non-idempotent.  They determine
the mentioned correlation entropy per particle $s=-\sum_\kappa
n_\kappa \ln n_\kappa /\sum_\kappa n_\kappa$. Knowledge of the PD
$n(x_{12})$ allows \cite{Sut71,RomS93,Ha96,For95} us to calculate the
fluctuation $\Delta N_X$ of the particle-number around its mean value
$N_X= n X$ in any piece (domain) $X$ of the $x$-axis.  Comparing this
variance of the particle-number distribution $P_X(N)$ for the cases
`no correlation' ($\nu=1/2$ or HF approximation) and `correlation'
shows the above mentioned narrowing for repulsion ($\nu>1/2$) in a
smaller $\Delta N_X$. Contrarily, for attraction ($\nu<1/2$) a
broadening with a larger $\Delta N_X$ appears.

In Section \ref{sec:system}, we introduce the CS model, define the
kinematical quantities used throughout the text, and present the
Hellmann-Feynman theorem. Section \ref{sec:thermo} is devoted to the
thermodynamic limit. In Section \ref{sec:correlation}, after
presenting the HF approximation, we discuss first qualitatively and
then analytically the influences of the CS interaction on 1PDM and PD.
In particular, we show that the above mentioned divergences in kinetic
and potential energies are caused by a peculiar behavior of the PD
$n(x_{ij})$ for {\em small inter-particle separations} $x_{ij}\ll
k^{-1}_{\rm F}$ and of the momentum distribution $n_\kappa$ for {\em
  large momenta} $k\gg k_{\rm F}$ or $\kappa \equiv k/k_{\rm F} \gg
1$.  In Section \ref{sec:corrmsr} we then apply the mentioned
fluctuation-correlation measures to the CS model.  Section
\ref{sec:numerics} is devoted to details of the numerics and in
Section \ref{sec:bosons} we discuss extensions of our approach to
impenetrable Bosons and lattice gases. We conclude in Section
\ref{sec:concl} with a discussion of our results. 

\section{The system and its ground state}
\label{sec:system}

\subsection{Hamiltonian, energies, and quantum kinematical quantities}
\label{sec:hamil}

The Hamiltonian of the CS model is $\hat{H} = \hat{T} + \hat{V}$ with
\begin{equation}\label{tv}
\hat{T} = \sum_{i}^{N}\frac{1}{2}p_i^2 \quad , \quad
\hat{V} = \sum_{i}^{N} v_{\rm ext}(x_i) +
 \sum_{i<j}^{N} \frac{\nu^2-\frac{1}{4}}{x_{ij}^2} \; ,
\label{eq:hamil}
\end{equation}
with $p_i^2 = - {{\partial}^2}/{\partial x_i^2}$, and $N$ equal to the
number of particles. We assume the system to be confined to the length
$L$ by an external potential $v_{\rm ext}(x)$, {\em e.g.}, a box or
harmonic oscillator potential. In the following we alternatively
assume periodic boundary conditions with $v_{\rm ext}(x)=0$ and a
density in the $k$ space described by $L\Delta k/(2\pi)=1$
\cite{Sut71,periodic}. The average particle density is $n = {N}/{L}$.
Furthermore, it follows from dimensional reasons that all energies for
the Hamiltonian (\ref{eq:hamil}) are proportional to $n^2$ in
agreement with the virial theorem and all lengths are measured in
units of $1/n$ (thus $k_{\rm F} \sim n$) \cite{Sut71}.

We denote the ground state energy and its kinetic and potential
`components' by $E_N = \langle \hat{H} \rangle $, $T_N = \langle
\hat{T} \rangle $, and $V_N = \langle \hat{V} \rangle $, respectively.
Then the corresponding energies per particle are $e_N = E_N/N$, $t_N =
T_N/N$, $v_N = V_N/N$ with $e_N = t_N + v_N$.
From the antisymmetric (or symmetric) ground state wave function
follow by contractions \cite{Sut71,RomS93} the 1PDM $\gamma _N(x; x')$
and the PD $n_N(x_1,x_2)$ normalized as $\int dx \gamma _N(x; x) = N$
and $\int dx_1 \int dx_2\ n_N(x_1,x_2) = N (N - 1)$, respectively.
The PD describes the XC hole, vanishing for zero separation and
approaching the Hartree product $\rho_N (x_1) \rho_N (x_2)$ for large
separations.  The cumulant PD $w_N(x_1,
x_2)\equiv\rho_N(x_1)\rho_N(x_2)-n_N(x_1,x_2)$ is via
\begin{equation}\label{eq:m}
  \int dx_1 \int dx_2\ w_N(x_1, x_2) = N
\end{equation}
size-extensively normalized.
Furthermore with the abbreviation $y=k_{\rm F}x$, $k_{\rm F}= \pi n$
\cite{Sut71}, and with the {\em dimensionless functions}
$f_N(y_1;y_2)$ hermitian, $g_N(y_1,y_2)$ non-negative, and
$h_N(y_1,y_2)\equiv f_N(y_1;y_1)f_N(y_2;y_2)-g_N(y_1,y_2)$, we can
write for the 1PDM $\gamma _N(x_1;x_2)=n\ f_N(y_1;y_2)$, for the PD
$n_N(x_1,x_2)= n^2\ g_N(y_1,y_2)$, and for the cumulant PD we have
$w_N(x_1,x_2)=n^2\ h_N(y_1,y_2)$.  The dimensionless cumulant PD is
thus $h_N=1-g_N$ and normalized as
\begin{equation}\label{eq:pdnorm}
{1\over N}\int {dy_1\over \pi}{dy_2\over
 \pi}\ h_N(y_1,y_2)=1,
\end{equation}
which follows from \Eref{eq:m}.  With these dimensionless 1PDM
and PD and with the Fermi energy $\epsilon_{\rm F}=k^2_{\rm F}/2$ the
energies $t_N$ and $v_N$ are given by
\numparts
\label{eq:tpv}
\begin{equation}\label{eq:t}
  t_N={1\over N} \int {dy_1\over \pi}
      {\left[ - \frac{\partial^2}{\partial y_1^2} f_{N}(y_1;y_2)\right]}_{y_2=y_1}
      \epsilon_{\rm F}
\end{equation}
and
\begin{equation}\label{eq:v}
v_N= {1\over N}
\int {dy_1\over \pi}{dy_2\over \pi}
g_N(y_1,y_2) {\mu\over y_{12}^2}
\epsilon_{\rm F}.
\end{equation}
\endnumparts
Therefore $t_N/\epsilon_{\rm F}$, $v_N/\epsilon_{\rm F}$ and
$e_N/\epsilon_{\rm F}$ are functions of $\mu$ and $N$. The latter
dependence disappears for the thermodynamic limit as shown in Section
\ref{sec:thermo}.

\subsection{The Hellmann-Feynman theorem}
\label{sec:rigorous}

If $e_N$ is known as a function of $\mu$, then $t_N$ and $v_N$ can be
obtained from the Hellmann-Feynman theorem \cite{Hel37} without
knowing the quantum-kinematical quantities $f_N(y_1;y_2)$ and
$g_N(y_1,y_2)$. This theorem says
\begin{equation}
\frac{\partial  E_N}{   \partial \mu }
= \biggl\langle   \frac{\partial \hat{H}}
{\partial \mu } \biggr\rangle
\label{eq:hellfeyn}
\end{equation}
which for (\ref{eq:hamil}) gives
\begin{equation}
v_N = \mu \frac{\partial e_N}{\partial \mu}, \quad
t_N = \left (
            1 - \mu \frac{\partial}{\partial \mu}
    \right ) e_N
\label{eq:hellfey-r2}
\end{equation}
and also
\begin{equation}\label{eq:lt}
 \frac{\partial}{\partial \mu}  t_N  =
 -\mu \frac{\partial}{\partial \mu}
  \left (
   \frac{1}{\mu} v_N
  \right ) \; .
\end{equation}
Thus --- with \Eref{eq:tpv} in mind --- the Hellmann-Feynman
relation (\ref{eq:hellfeyn}) for the $1/{x_{ij}^2}$ model establishes
an integral relation between the dimensionless 1PDM $f_N$ on the
l.h.s.\ and the dimensionless PD $g_N$ on the r.h.s.\ of \Eref{eq:lt}.
%

\section{Thermodynamic limit}
\label{sec:thermo}

We wish to study the thermodynamic limit with $N \rightarrow\infty$
and $L \rightarrow \infty$ such that $n=N/L = \mbox{const}$. The
resulting extended system has only two parameters, the
interaction strength parameter $\nu$ and the Fermi
wave number $k_{\rm F}$.  So $t/\epsilon_{\rm F}$,
$v/\epsilon_{\rm F}$, and $e/\epsilon_{\rm F}$ become functions of
$\nu$ only.
The thermodynamic limit makes furthermore the 1PDM and the PD to
depend only on $k_{\rm F}x_{12}\equiv k_{\rm F}(x_{1}-x_{2})= y_1-y_2
\equiv y_{12}$. The dimensionless functions $f_N$, $g_N$, and $h_N$
then take the forms $f(y_{12})$, $g(y_{12})$, and
$h(y_{12})=1-g(y_{12})$, respectively, with $f(0)=1$ and $g(0)=0$ or
equivalently $h(0)=1$.  These functions have $\nu$ as the only
parameter.

The eigenfunctions (or natural orbitals) of the 1PDM $\gamma (x_{12})=n
f(y_{12})$ become simply plane waves $ {\varphi}^0_k(x) = {\rm e}^{{\rm
    i}kx}/L$, such that
\begin{equation}
\label{eq:16}
  \gamma (x_{12})
    =  n\ \int_0^\infty d\kappa \ n_{\kappa}
           \cos \kappa |y_{12}|
    \equiv  n\ f(y_{12}) ,
\end{equation}
where $n_\kappa$ is the momentum distribution and $\kappa =k/k_{\rm F}$.

For $\nu=1/2$ (ideal spinless 1D Fermi gas) the Pauli principle leads
in the reciprocal space to the Fermi ice block
$n^0_\kappa=\theta(1-|\kappa|)$ and in the direct space to the ideal X
hole $g^{0}(y)=1-{[f^{0}(y)]}^{2}$ with the dimensionless 1PDM
$f^0(y)=(\sin y)/y$ following from \Eref{eq:16}. The energy per
particle is $e_0=\epsilon_{\rm F}/3= k^2_{\rm F}/6$.

In general, with $\gamma (0) = n$, the momentum distribution
$n_{\kappa}$ is normalized as $\sum_{\kappa} n_{\kappa} = N$ or
\begin{equation}\label{eq:nk}
\int_{0}^{\infty}  d\kappa \  n_{\kappa } = 1 \; .
\end{equation}
The kinetic energy per particle is according to \Eref{eq:t}
\begin{equation}\label{eq:tk}
  t
  = 6 \int_{0}^{\infty}  d\kappa \ n_{\kappa } \frac{\kappa ^2}{2}   e_0 \; .
\end{equation}
$n_{\kappa }$ is a function of ${|\kappa |}$ and $\nu$, so $t/e_0$ is
a function of $\nu$ only with $t=e_0$ for $\nu =1/2$.

The corresponding expressions for the PD $g(y)$ are according to
\Eref{eq:pdnorm}
\begin{equation}\label{eq:pdd}
 2 \int_{0}^{\infty}
     \frac{dy}{\pi} \ h(y)  = 1 , \; h(y)= 1 - g(y)
\end{equation}
and for the interaction energy per particle according to \Eref{eq:v}
\begin{equation}\label{eq:ie}
v
  = 6 \int_{0}^{\infty}
               \frac{dy}{\pi}\  g(y) \frac{\nu^2+\frac{1}{4}}{ y^2}  e_0
 \; .
\end{equation}
$g(y)$ is a function of $y$ and $\nu$, so $v/e_0$ is a function of
$\nu$ only. With $t$ and $v$ follows the integral relation
\begin{equation}\label{eq:zz}
\int\limits^\infty_0d\kappa\
 \frac{\kappa^2}{2}
 \frac{\partial n_\kappa}{\partial \mu}
=
-\int\limits^\infty_0\frac{dy}{\pi}\
 \frac{\nu^2+\frac{1}{4}}{y^2}
 \frac{\partial g(y)}{\partial \mu}
\end{equation}
as a consequence of the Hellmann-Feynman theorem expressed in
\Eref{eq:lt}. Correlation via $\nu\neq 1/2$ deforms the X hole and the
Fermi ice block as shown in Figs.\ \ref{fig:PD} and
\ref{fig:nk-scaled-CF} in such a way that \Eref{eq:zz} is
maintained.

\section{Hartree-Fock approximation and correlation beyond it}
\label{sec:correlation}

\subsection{Hartree-Fock (HF) approximation}
\label{sec:hf}

The simplest approximation for the quantities $n_\kappa$, $g(y) $,
$t$, and $v$ is obtained from the HF approach. The momentum
distribution in \Eref{eq:tk} and the PD in \Eref{eq:ie}
are to be replaced by their `ideal' expressions $n^0_\kappa $ and
$g^0(y)$, respectively.  Consequently, we find $t_{\rm HF} = e_0$ and
$v_{\rm HF} = 2\mu e_0$ and thus $ e_{\rm HF} = ( 1 + 2 \mu ) \; e_0$,
as shown in \Fref{fig:Eng-BAvsHF}.  Here the identity
(\ref{eq:a25}) has been used.  The total HF energy $e_{\rm HF}$ also
obeys the Hellmann-Feynman theorem (\ref{eq:hellfeyn}) and the virial
theorem.

\subsection{Qualitative discussion of correlation}
\label{sec:general}

Due to correlation the true ground state energy per particle, $e$, is
below the HF energy $e_{\rm HF}$ and the true ground state wave
function $\Phi (\cdots)$ is no longer a single Slater determinant.
Note that the definition of the term `correlation' needs a reasonable
reference state, which is ${\Phi}_{\rm HF} (\cdots)$ in our case. So,
correlation causes a negative correlation energy $e_{\rm corr} = e -
e_{\rm HF} < 0$, namely through redistributions of $g^0(y)$ and
$n_\kappa ^0$ which are shown in Figs.\ \ref{fig:PD} and
\ref{fig:nk-scaled-CF} and described in the following.

\begin{figure}[p]
\centerline{\psfig{figure=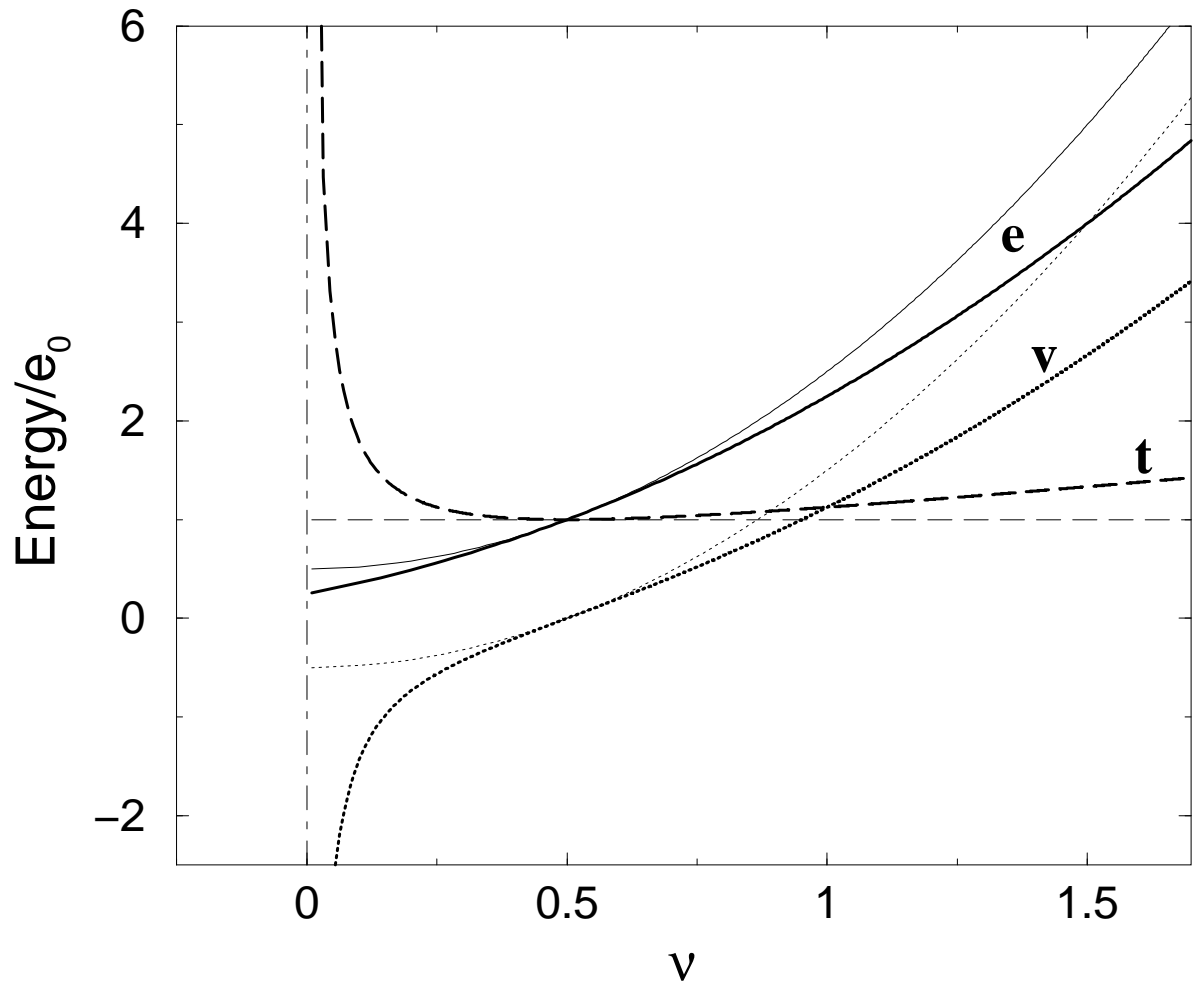,width=\columnwidth}}
\caption{\label{fig:Eng-BAvsHF}
  Bulk energy $e$ (solid), kinetic energy $t$ (dashed), and potential
  energy $v$ (dotted) plotted as functions of interaction strength
  parameter $\nu$.  Thin lines denote the results of the Hartree-Fock
  approximation, thick lines are exact. The thin dashed-dotted line
  indicates the ``fall-into-the-origin'' at $\nu=0$.}
\end{figure}

As we show in \Fref{fig:PD}, correlation modifies the X hole of
the unperturbed PD. Especially the correlation induced changes for
small $y$ are of interest, because the interaction $1/y^2$ is there
largest.  The on-top behavior of the uncorrelated X hole ($\nu=1/2$ or
HF) is described by $g^0(y)=y^2/3+\ldots$.  In its correlated
counterpart with a $\nu$-dependent exponent and $\nu$-dependent
coefficients (see \ref{sec:kimball})
\begin{eqnarray}\label{eq:kimball}
g(y) &= & A {  y }^ {\alpha }\left( 1+a_1y+a_2y^2+\cdots \right) , \quad
\nonumber \\
\alpha
&= &1+ 2\nu, 
\end{eqnarray}
correlation for $\nu \neq 1/2$ shows up in $\alpha\neq 2$ and $A\neq
1/3$. More precisely, repulsive particle interaction $(\nu > 1/2)$
supports the Pauli `repulsion', so the X hole is broadened (through
increasing $\alpha$ and decreasing $A$), but attractive particle
interaction $(\nu < 1/2)$ fights against (or competes with) the Pauli
`repulsion', so the X hole is narrowed (through decreasing $\alpha$
and increasing $A$) as shown in \Fref{fig:PD} and \Tref{tab-2}.  This
X hole narrowing (for $\nu < 1/2$) or broadening (for $\nu > 1/2$)
makes
\begin{equation}
\label{eq:denom}
6\int^\infty_0 \frac{dy}{\pi}\ \frac{g(y)}{y^2}=
1+\frac{1}{2\nu} \gtrless 2 \quad {\rm for}\quad
\nu \lessgtr \frac{1}{2}.
\end{equation}
The equation follows from \Eref{eq:ie} together with the
Hellmann-Feynman theorem (\ref{eq:hellfeyn}). Thus $v {<} v_{\rm HF}=
2\mu e_0$ for $ \nu \neq 1/2$ as shown in \Fref{fig:Eng-BAvsHF}.

As shown in \Fref{fig:nk-scaled-CF}, correlation thaws the Fermi ice
block $\theta(1-|\kappa|)$. Mathematically, the momentum distribution
$n_\kappa$ becomes non-idempotent, physically this means: Correlation
excites particles for $\kappa >1$ and holes for $\kappa<1$. This
increases the kinetic energy independent whether the interaction is
attractive $(\nu < 1/2)$ or repulsive $(\nu > 1/2)$: $t > t_{\rm HF}$
as can be seen in \Fref{fig:Eng-BAvsHF}.  We note that $n_\kappa$ has
no discontinuity at $|\kappa| = 1$.  Its value is ${1}/{2}$ and near
$\kappa=1$ it follows a power law as is typical for all Luttinger
liquids with their $z_{\rm F}=0$ \cite{Hal81b,SchCP98}.

We model the melting of the Fermi ice block analytically with
the continuous function
\numparts
\begin{eqnarray}
\label{eq:nk01}
n_\kappa & = &  {1\over 2}+B(1-\kappa)^\beta
\left[ 1+b^-_1(1-\kappa)^\beta + b^-_2(1-\kappa)^{2\beta}+\ldots\right]
\nonumber \\ & & \quad {\rm for\ } 0\leq\kappa\leq 1, \\
\label{eq:nk12}
n_\kappa & = &  {1\over 2}-B(\kappa-1)^\beta
\left[ 1+b^+_1(\kappa-1)^\beta+b^+_2(\kappa-1)^{2\beta}+\ldots\right]
\nonumber \\ & & \quad {\rm for\ } 1\leq\kappa\leq 2, \\
\label{eq:nk2i}
n_\kappa & = &  {C\over {\kappa^\gamma}}
\left( 1+{c_2\over \kappa^2}+{c_4\over \kappa^4}+\ldots \right)
\nonumber \\ & & \quad {\rm for\ } 2\leq\kappa\leq \infty
\end{eqnarray}
\label{eq:nkall}
\endnumparts
with \cite{RomS93,KawY91}
\begin{equation}
\beta = {1\over 4}\frac{(1-2\nu)^2}{1+2\nu}
\label{eq:beta}
\end{equation}
and $\gamma=3 + 2\nu$ (\ref{sec:kimball}). The exponents $\beta,\gamma$
and the coefficients $B,C$, and $b^{\pm}_i,c_i$ depend on $\nu$. The
condition $0<\beta<1$ ensures an infinite slope of $n_{\kappa}$ at
$\kappa=1$. This $n_\kappa$ has to obey the
normalization (\ref{eq:nk}) and the condition
\begin{equation}
  3 \int^\infty_0 d\kappa \ n_\kappa \kappa^2 = \frac{\left({1\over
      2}+\nu\right)^2}{2\nu} \geq 1 \ ,
\end{equation}
which follows from \Eref{eq:tk} together with the
Hellmann-Feynman theorem (\ref{eq:hellfeyn}). For $\nu=1/2$ (or HF) it
is $\beta=0$, $B(1+\sum_i b^{\pm}_i)={1\over 2}$, and $C=0$.  The
correlation induced melting for $\nu\neq 1/2$ shows up in $\beta>0$,
$B(1+\sum_i b^{\pm}_i)<{1\over 2}$, and $C > 0$.

\subsection{Results of the exact solution}
\label{sec:bethe}

With the help of the Bethe-Ansatz technique one obtains $e =
{\lambda}^2 e_0$ \cite{Sut71,RomS93}.  $e$ as a function of the
interaction strength parameter $\lambda$ shows no special behavior for
$\lambda {> \atop \to} 1/2$, but as a function of the interaction
strength $\mu=\lambda(\lambda -1)$,
\begin{equation}\label{eq:ee}
e= \left({1\over 2}+\nu \right)^2 e_0 , \quad \nu =\sqrt{{1\over 4}+\mu}
\end{equation}
the non-analytical behavior for $\mu \rightarrow -{1/4}$ is
incorporated in the variable $\nu$.

\Eref{eq:ee} yields with the Hellmann-Feynman theorem
(\ref{eq:hellfey-r2}) the kinetic energy per particle,
\begin{equation}\label{eq:gll}
t= \frac{\left({1\over 2}+\nu \right)^2}{2\nu} e_0.
\end{equation}
\Eref{eq:ee} yields with \Eref{eq:hellfey-r2}
also the interaction energy per particle
\begin{equation}\label{eq:hll}
v=\mu \left(1+\frac{1}{2\nu}\right) e_0.
\end{equation}
From \Fref{fig:Eng-BAvsHF} we see that both $t$ and $v$ diverge for
$\nu {\to} 0$, while $e$ remains finite.

Equations  (\ref{eq:gll}) and (\ref{eq:hll}) lead to
\begin{equation}\label{eq:inr}
\int_{0}^{\infty}  d\kappa \ n_\kappa  \kappa^2
     = 6\nu\left[
       \int_{0}^{\infty}
       \frac{dy}{\pi}\   \frac{g(y)}{y^2 }\right]^2  ,
\end{equation}
as another integral relation between the momentum distribution
$n_{\kappa}$ and the dimensionless PD $g(y)$ in addition to
\Eref{eq:zz}. These distribution functions have to change with $\nu$
in such a way that these relations (\ref{eq:zz}) and (\ref{eq:inr})
are obeyed together with the normalization conditions (\ref{eq:nk})
and (\ref{eq:pdd}).

The PD $n(x_{12})=n^2 g(y)$ with $y=k_{\rm F}x_{12}$ is known
analytically for the values $\nu = 0$, ${1}/{2}$, and
${3}/{2}$ \cite{Sut71,RomS93}. For $\nu = 0$ it is (with the notation
of  \ref{sec:integrals})
\begin{equation}\label{eq:nx12}
g(y)= 1 - {\left ( \frac {\sin y}{y} \right )}^2
        + {\rm Si}(y) { \frac {d}{dy} \frac {\sin y}{y}}
        - \frac {\pi}{2} \frac {d}{dy} \frac {\sin y}{y} ,
\end{equation}
for $\nu ={1}/{2}$ (ideal Fermi gas) it is
\begin{equation} \label{eq:g}
g(y)= 1- {\left ( \frac{\sin y}{y} \right )}^2  ,
\end{equation}
and for $\nu = {3}/{2}$ it is
\begin{equation}\label{eq:nxl}
g(y)=1 - {\left ( \frac{\sin 2y}{2y}\right )}^2
+ {\rm {Si}}(2y) \frac{d}{d(2y)}   \frac{\sin 2y}{2y} .
\end{equation}
The corresponding dimensionless cumulant PDs $h(y)=1-g(y)$ are given
in \Tref{tab-1} together with their Fourier transforms
\begin{equation}\label{eq:fou}
{\tilde h}(q)=2\int^\infty_0 dy \cos (qy)\; h(y).
\end{equation}
They have via $S(q)=1-{\tilde h}(q)/\pi$ a simple relation to the
static structure factor (or van Hove correlation function) $S(q)=
\langle{\hat \rho}_{q}{\hat \rho}^\dag_{q}\rangle/N$, which describes the
correlation of density-density fluctuations.  ${\hat \rho}_{q}=\sum_i
{\rm exp}(-{\rm i}{q}{x}_i)$ is the Fourier transform of the density
operator ${\hat \rho}({x})= \sum_i \delta({x}-{x}_i)$. The three PDs
$g(y)$ are shown in \Fref{fig:PD} and the three structure factors
$S(q)$ in \Fref{fig:Sq}.

For $\nu=1/2$ the weak oscillations of $g(y)$ and the (first-order) kink
of $S(q)$ arise from the Fermi momentum distribution $n_{\kappa}$ with
its sharp discontinuity $z_{\rm F}=1$ at $\kappa=1$. With increasing
repulsion the oscillations of $g(y)$ are enhanced, what is displayed in
the reciprocal space by the peak of $S(q)$ at $q=2$ (and a 3rd-order
kink at $q=4$). This peak is customary in 1D quantum liquids, see, e.g.,
the spin-correlation functions of Ref.\ \cite{YokO91}.  The first
maximum of $g(y)$ runs through a certain trajectory from ($\pi$, $1$) to
($2.99$, $1.24$).  Whereas repulsion enhances the Friedel oscillations
of $g(y)$ and the kink structure of $S(q)$, they diminish with
increasing attraction: for $\nu=0$ both $g(y)$ and $S(q)$ approach the
value $1$ smoothly (non-oscillatory) from below. The discontinuity for
$\nu=0$ of the second derivative of $S(q)$ at $q=2$ replaces the kink
seen for other values of $\nu$.
For the on-top behavior of $g(y)$ in terms of $g(0)$, $g'(0)$,
$g''(0)$ the following holds: It is $g(0)=0$, according to the Pauli
principle, $g'(0)=\pi/6$ for $\nu=0$, but $0$ for $\nu>0$, and it is
$g''(0)=0$ for $\nu=0$, infinite for $0<\nu<1/2$, but $2/3$ for
$\nu=1/2$, and $0$ for $\nu >1/2$. We remark that $S(q)$ and $g(y)$
can be computed also for other values of $\nu$ in addition to the
three special values used here \cite{Ha96,For95}.

With the identities (\ref{eq:a22})--(\ref{eq:a24}) the normalization
condition (\ref{eq:pdd}) is fulfilled.  From Equations
(\ref{eq:nx12})--(\ref{eq:nxl}) follow the on-top coefficients of
\Eref{eq:kimball}; they are shown in \Tref{tab-2}.  Note that the
last two terms of \Eref{eq:nx12} do not contribute to the
normalization because of \Eref{eq:a23} and that the last term
causes the odd on-top coefficients of \Tref{tab-2} and also the
linear behavior for small $y$. Its oscillations are exactly canceled
by the combined oscillations of the second and the third term.
Simultaneously, these terms ensure the correct normalization.

The PD (\ref{eq:g}) for $\nu\rightarrow 1/2$ plugged into \Eref{eq:v}
yields with the identity (\ref{eq:a25}) the same as results from
\Eref{eq:hll}, which follows from the total energy per particle,
\Eref{eq:ee}, and the Hellmann-Feynman theorem (\ref{eq:hellfey-r2}), namely
$v=2 \mu e_0$. Similarly the PD (\ref{eq:nxl}) for $\nu \to 3/2$
plugged into \Eref{eq:ie} yields with the identities (\ref{eq:a25})
and (\ref{eq:a26}) the same as results from \Eref{eq:hll}), namely $v=
{4}\mu e_0/{3} $.

For $\nu = 0$ a divergence appears, because from the PD
(\ref{eq:nx12}) follows an on-top behavior, which is linear in $y$ as
shown in \Fref{fig:PD} and \Tref{tab-2}.  This linear
behavior results from the last term of \Eref{eq:nx12}, which does not
influence the normalization (\ref{eq:pdd}), but it makes the
interaction energy $v \sim \int_{0}^{\infty} dy \ g(y)/{y}^2$ to
diverge logarithmically in agreement with the divergence of $v
\rightarrow - {e_0}/{8}{\nu}$ for $\; \nu {> \atop \rightarrow} 0 $ as
displayed in \Fref{fig:Eng-BAvsHF}.

The divergence of the interaction energy is accompanied and
compensated by the corresponding divergence of the kinetic energy $t
\rightarrow e_0/8\nu$. This indicates a special asymptotic behavior of
the momentum distribution $n_\kappa $ for $\nu {> \atop \rightarrow}
0$, namely \Eref{eq:nk2i} with $\gamma {> \atop \rightarrow} 3$. For
$\gamma > 3$ the integral $\int_{0}^{\infty} d\kappa \ n_\kappa
\kappa^2$ is convergent, but with $\gamma {> \atop \rightarrow} 3$ for
$\nu {> \atop \rightarrow} 0$ it diverges logarithmically, whereas the
normalization integral (\ref{eq:nk}) remains convergent. The
counterpart to this asymptotic behavior of $n_\kappa$ for $\kappa
\rightarrow \infty$ is the on-top behavior of the PD for $y
\rightarrow 0$ as shown in \Fref{fig:PD} and \Tref{tab-2}
with a smooth transition of the coefficient $A$ in \Eref{eq:kimball}
from $\pi /6$ via $1/3$, to $16/135$ for $\nu=0,\ 1/2$ and $3/2$,
respectively.  With quadratic interpolation of the coefficients shown
in \Tref{tab-2} as functions of $\nu$, one may continuously
switch the on-top behavior of the PD $g(y)$ between its forms at
$\nu=0$ and $3/2$. For the PD exponent $\alpha =1 + 2\nu$ we refer to
 \ref{sec:kimball}, where also the momentum-distribution
exponent is conjectured as $\gamma=3+2\nu$.

These divergences of the kinetic and the interaction energies indicate
that for attractive particle interaction $\mu /{x_{ij}^2}$ with $\mu
{\to} -1/4$ the system becomes unstable (no ground state with finite
kinetic and potential energies). We remark that it was shown in Ref.\
\cite{LanL77} that the singular particle interaction $ - {|\mu|
  }/{|\vec{r}_{12}|^2}$ makes already two particles to fall together
(``fall-into-the-origin'') for $|\mu| > {1}/{4}$ (ground state with $E
\rightarrow - \infty$) and for $|\mu| < {1}/{4}$ there are only
scattering states with $E \ge 0$ (no bound states with $E <
0$) \cite{Sut71}.

For $\nu =0$ the exact solution of the CS model yields the
momentum-distribution data. In Section \ref{sec:numerics} we will give
the details of the necessary numerical calculation. The coefficients
of \Eref{eq:nk01} are fitted to the $n_\kappa$ values for
$\kappa=0\ldots 1$ and the coefficients of \Eref{eq:nk12} are
fitted to the $n_\kappa$ values for $\kappa =1\ldots 2$. The
coefficients are chosen to make also $n_\kappa$ at $\kappa=2$
continuous and smooth.  Finally $b^+_3$ is fine tuned to make the
normalization equal to $1$ according to (\ref{eq:nk}). The results are
shown in \Fref{fig:nk-scaled-CF} and the values of the
coefficients are given in \Tref{tab-4p5}.
The case $\nu=3/2$ is similarly treated only with the difference that
also the kinetic energy $t={4\over 3}e_0$ can be used for the fine
tuning in addition to the normalization condition (\ref{eq:nk}). The
results are shown also in \Tref{tab-4p5}.  Here $b^+_3$ and $b^+_4$
have been used for fine tuning (which yields the normalization $0.997$
and $t=1.34\ e_0$, instead of 1 and $4e_0/3$).

\section{Fluctuation-correlation analysis}
\label{sec:corrmsr}

\subsection{Quantities following from the pair density}
\label{sec:corrmsr-pairdens}

\paragraph*{Particle-number fluctuations:}
Following Fulde \cite{Ful95} one may ask to what extent correlation
influences particle-number fluctuations $\Delta N_X$ in a domain $X$,
{\em i.e.}, a certain interval of the $x$ axis, where in the average
there are $N_X=n X$ particles. These fluctuations are measured
quantitatively by \cite{Ful95,ZieTSP99,Zie00a}
\begin{eqnarray}
\frac{\left(\Delta N_X\right)^2}{N_X}
&=&1-\frac{1}{n X}
\int^X_0 dx_1 \int^X_0 dx_2\ w(x_{12}) \nonumber \\
&=&1-{1\over Y}\int^Y_0 dy_1 \int^Y_0 dy_2\ \frac{h(y_{12})}{\pi},
\label{eq:partnumb}
\end{eqnarray}
with $Y=k_{\rm F}X=\pi n X$. Following the Appendix A of Ref.\
\cite{ZieTSP99} the 2D integral (\ref{eq:partnumb}) is reduced to a 1D
integral with the help of the Fourier transform (\ref{eq:fou}), namely
\begin{equation}\label{eq:a42}
\frac{\left(\Delta N_X\right)^2}{N_X}=1-\frac{2}{Y}\int^\infty_0
{dq\over \pi} \frac{1-\cos (qY)}{q^2}\; \frac{{\tilde h}(q)}{\pi}.
\end{equation}
The results are shown in \Fref{fig:part-num-fluc}, where also the case
$\nu \rightarrow \infty$ (`strict' or `perfect' correlation)
\cite{ZieTSP99} is displayed. Traces of the oscillations as
$\nu\rightarrow\infty$ are already visible for $\nu=3/2$.

With $h(y)=1-g(y)$ and with the expansion of $g(y)$ according to
\Eref{eq:kimball} --- see also the text after \Eref{eq:nxl} --- the
small-$X$ expansion of \Eref{eq:partnumb} is
\begin{eqnarray}
\frac{\left(\Delta N_X\right)^2}{N_X}
& = & 1 + d_1 n X
+ d_2 (n X)^2
+ d_3 (n X)^3 \nonumber \\
& & \mbox{ }
+ d_4 (n X)^4
+ d_5 (n X)^5
+ \ldots \quad .
\label{eq:fluccoeff}
\end{eqnarray}
The slope $d_1$ at $X=0$ does not depend on the interaction strength
parameter $\nu$ as shown in \Tref{tab-3} because of $g(0)=0$ and
$h(0)=1$ not depending on $\nu$; but the coefficients of the next
terms do such that the particle-number fluctuations are suppressed due
to repulsive particle interaction, but enhanced due to attractive
particle interaction: correlation makes the particle-number
distribution $P_X(N)$ more narrow for repulsion ($\nu>1/2$) and more
broad for attraction ($\nu<1/2$). We remark, that fluctuation
enhancement (induced by attractive interaction) generally may
support/cause clusterings ({\em e.g.}, paramagnons prior the
para-to-ferromagnetic phase transition). In our case this tendency
shows up in the sudden ``fall-into-the-origin'' at $\nu=0$.
If one considers with $X=1/n$ a Wigner-Seitz `sphere'
(with `radius' $X/2$ and $N_X=1$), then
\begin{equation}\label{eq:fluc}\label{eq:sig1}
\Sigma_1(\nu)=1-\frac{\chi(\nu)}{\chi(1/2)}, \quad
\chi(\nu)=\frac{\left(\Delta N_X\right)^2}{N_X}
\end{equation}
is a reasonable correlation measure based on particle-number
fluctuations as we show in \Fref{fig:corr-msr-all}.

\paragraph*{On-top behavior:}
The exponent $\alpha$ and the coefficients $A, a_i$ of
\Eref{eq:kimball} describe the short-range or dynamical correlation,
{\em i.e.}, the small-separation behavior of $g(y)$, see \Tref{tab-2}.
Cioslowski's correlation cage \cite{CioL99} is in our case simply the
inter-particle-separation range $y=0\ldots y_{\rm max}$ with $y_{\rm
  max}$ being that separation where the PD $g(y)$ has its first
maximum $g_{\rm max}=g(y_{\rm max})$. For $\nu =0,1/2,3/2$ the
corresponding values are $y_{\rm max}=\infty $, $\pi$, $2.99$ and
$g_{\rm max}$= $1$, $1$, $1.24$ \cite{ZieTSP99}.  One may ask to what
extent the correlation cage contributes to the interaction energy and
define
\begin{equation}
\label{eq:sig2}
\Sigma_2(\nu)=1-\frac{V_{\rm cage}(\nu)}{V_{\rm cage}(1/2)},
V_{\rm cage}(\nu ) =
\frac{ \int^{y_{\rm max}}_0 {dy} \ g(y)/{y^2}}
                   {\int^{\infty}_0 {dy} \ g(y)/{y^2}}\leq 1
\label{eq:Vcage}
\end{equation}
as an energetic correlation measure with $V_{\rm cage}(0)=1$; the
expression simplifies when using (\ref{eq:denom}).  Both $\Sigma_1$
and $\Sigma_2$ vanish for $\nu=1/2$ as shown in \Fref{fig:corr-msr-all}.

\subsection{Quantities following from the momentum distribution}
\label{sec:corrmsr-momdist}

\paragraph*{Critical exponent:}
The critical or correlation exponent $\beta$ of \Eref{eq:beta} can be
computed from conformal field theory \cite{RomS93,KawY91}. It
describes (together with the coefficient $B$) the behavior of
$n_\kappa$ near $\kappa=1$ according to Equations (\ref{eq:nk01}) and
(\ref{eq:nk12}). For the three special values $\nu=0,1/2$ and $3/2$,
this gives $1/4, 0$, and $1/4$, respectively, as shown in
\Fref{fig:corr-msr-all}.  The exponent $\gamma$ describes the decay of
the correlation tail.

\paragraph*{Non-idempotency and correlation entropy:}
The $q$th order non-idempotency is \cite{ZieGJB97} $c(q)=1-\int^\infty_0
d\kappa \ (n_\kappa)^q$.  The derivative of $c(q)$ at $q=1$ is
$s\equiv c'(1)$ or
\begin{equation}
s(\nu )=-\int^\infty_0 d\kappa \ n_\kappa \ln n_\kappa \geq 0
\label{eq:corr-ent}
\end{equation}
to be referred to as correlation entropy \cite{ZieGJB97,Zie00a}. It
has been plotted in \Fref{fig:corr-msr-all}.

\paragraph*{Correlation tail properties:}
The relative number of particles (or holes) in the
corresponding correlation tail is \cite{ZieGJB97,Zie00a,TakY91}
\begin{equation}
N_{\rm tail}(\nu )=\int^\infty_1 d\kappa \
n_\kappa= \int^1_0 d\kappa \ (1-n_\kappa)<1.
\label{eq:Ntail}
\end{equation}
The contribution of the correlation tail to $s$ is
\cite{Zie00a}
\begin{equation} \label{eq:tail-entropy}
S_{\rm tail}(\nu )=-\int_{1}^{\infty} d\kappa \ n_{\kappa} \ln
n_{\kappa} < s(\nu).
\label{eq:Stail}
\end{equation}
In addition to these quantum-kinematic measures one may use
\cite{Zie00a}
\begin{equation}
T_{\rm tail}(\nu )= \frac{\int^\infty_1 d\kappa \ n_\kappa \kappa^2}
{\int^\infty_0 d\kappa \ n_\kappa \kappa^2} \leq 1
\label{eq:Ttail}
\end{equation}
as another energetic measure with $T_{\rm tail}(0)= 1$.  Also these
correlation measures vanish for $\nu=1/2$ as shown in \Fref{fig:corr-msr-all}.

\subsection{The correlation energy}
\label{sec:corrmsr-correng}

For $e_{\rm corr}=e-e_{\rm HF}$ follows
\begin{equation}\label{fc}
e_{\rm corr}= -\left (\nu -{1\over 2}\right )^2 e_0\;.
\end{equation}
Kinetic and interaction energy contribute $t_{\rm corr}=
-\frac{1}{2\nu} e_{\rm corr} $ and $ v_{\rm corr}= \left( 1 + {1\over
    2\nu}\right ) {e_{\rm corr}}$ , respectively, to $e_{\rm corr}$.

\subsection{Comparison of the correlation measures}
\label{sec:corrmsr-discussion}

When comparing the computed correlation measures in Figs.\ 
\ref{fig:corr-msr-all} it turns out that for small $|\nu -1/2|$ the PD
based measures $\Sigma_{1,2}$ of Eqs.  (\ref{eq:fluc}) and
(\ref{eq:Vcage}) are proportional to $\nu-1/2$ (which is $-e'_{\rm
  corr}/(2e_0)$), whereas the $n_\kappa$ based measures
(\ref{eq:corr-ent})--(\ref{eq:Ttail}) as well as (\ref{eq:beta})
behave like $(\nu-1/2)^2$ (which is $-e_{\rm corr}/e_0$).  So the
latter ones are not so sensitive as the first ones. With $s(\nu)=
0.5828 |e_{\rm corr}/e_0|+\ldots$ the Collins' conjecture $|e_{\rm
  corr}|\sim s$ is confirmed at least for weak interaction. In this
limit also $N_{\rm tail}$, $S_{\rm tail}$ and $T_{\rm tail}$ are
mutually proportional and their derivatives are proportional to
$\Sigma_{1}$ and $\Sigma_{2}$.

We remark that the quantities $\chi(\nu)$, $V_{\rm cage}(\nu)$, $N_{\rm
tail}(\nu)$, $S_{\rm tail}(\nu)$, and $T_{\rm tail}(\nu)$ are reference
free, {\em i.e.}, they are defined without reference to the
non-interacting case $\nu=1/2$ --- which in our case is simultaneously
equivalent to the Hartree-Fock approximation.  Reference quantities
appear in $\Sigma_{1,2}$ with $\chi(1/2)$ and $V_{\rm cage}(1/2)$ and in
$s(\nu)$ with $s(1/2)=0$.  Whereas this observation is important for
quantum chemistry --- as stressed by J.\ Cioslowski \cite{CioL99} ---
whenever multi configuration appears, it is less important in our case
which is well described by single configuration.

\section{Numerical determination of 1PDM and $\lowercase{n(\kappa)}$ for the CS model}
\label{sec:numerics}

Results from the theory of random matrices enable the calculation of
various correlation functions for the CS
model \cite{Sut71,RomS93,SimA93a,SimA93b,KraT00}. In particular, the
1PDM can be expressed in terms of a determinant of an appropriate
matrix $F^{(\nu)}_{pq}$ \cite{Sut71,RomS93}.  The size of this matrix
is specified by the number of particles $N$ to be $(N-1)^2$ for $\nu=
1/2$ and $3/2$ and $(N-1)^2/4$ for $\nu=0$.  Each element of
$F^{(\nu)}_{pq}$ contains simple trigonometric 1D ($\nu=1/2$ and
$3/2$) or 2D ($\nu=0$) integrals.

\begin{figure}[p]
\centerline{\psfig{figure=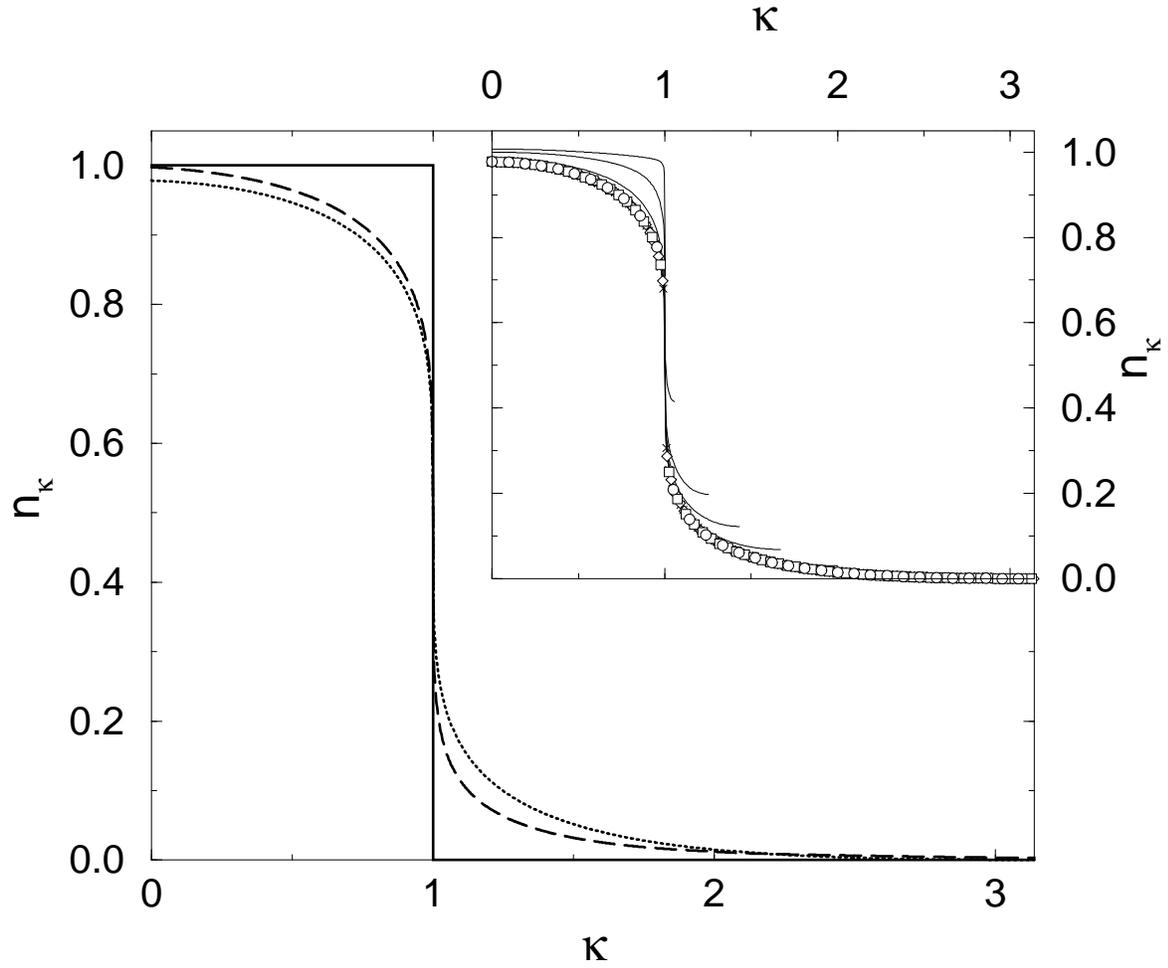,width=\columnwidth}}
   \caption{\label{fig:nk-scaled-CF}
     Fermionic momentum distributions $n_{\kappa}$ vs.\ $\kappa =
     k/k_{\rm F}$ with $\nu=0$ (dashed), $1/2$ (solid), and $3/2$
     (dotted). Inset: $n_{\kappa}$ for $\nu= 3/2$ and $L=401$ computed
     with $N= 21$, $41$, $81$, $121$, $161$, $201$, $241$, $281$,
     $321$, $361$, and $401$. The data for $N= 21 (\circ)$,
     $41(\square)$, $81(\diamond)$, and $121(\times)$ do not show any
     density dependence whereas the larger density data (lines) do.}
\end{figure}

For some cases, most notably $\nu=1/2$, the resulting determinant can
be computed analytically and corresponding expressions have been given
in Ref.\ \cite{RomS93}. For the other cases, we have evaluated the
determinant numerically \cite{RomS93}, using a subdivision of the
system volume (periodicity length) according to $L/L_0= 42, 402$, and
$402$ for $\nu= 0$, $1/2$, and $3/2$, respectively. The particle
number, odd due to periodicity of the wave function \cite{RomS93},
varied from $N=1$ to $401$, corresponding to a variation in density
$n$ from nearly $0$ to nearly $1$. Taking the Fourier transform, we
next compute the momentum distribution $n_\kappa$ for all densities.
In the inset of \Fref{fig:nk-scaled-CF}, we show results for one of
the three special $\nu$ values.

Next, we apply the definitions of correlation measures and correlation
energies as given in Sections \ref{sec:thermo}, \ref{sec:correlation},
and \ref{sec:corrmsr} and study their density dependence. In
\Fref{fig:cb2-cf2-CE} we show results for the entropy $s$ and in
\Fref{fig:cb2-cf2-eng} for the various energies as the density is
varied.  As {\em all} energies scale with $n^2$, these measures should
be {\em density independent} when normalized with respect to $e_0$.
However, we do in fact see a pronounced density dependence for $n
\gtrsim 0.5/L_0$ and also for $n \lesssim 0.05/L_0$. This latter
density dependence is simply due to the small particle numbers, thus a
small size of $F^{(\nu)}_{pq}$ and consequently a limited resolution
when computing the 1PDM at fixed $L/L_0$. The density dependence at
large $n$ values is more intricate to explain. The computation of the
1PDM by the connection with random matrix theory works for the {\em
  periodic} model \cite{periodic}. Thus there exists a Brillouin zone
and the tail of $n_\kappa$ for $|\kappa|$ outside this Brillouin zone
is folded back into it. The tail of $n_\kappa$ thus tends to be
dominated by this effect for large $n$ values as shown in the inset of
\Fref{fig:nk-scaled-CF}.  However, knowing that the correlation
measures must be independent of density in the thermodynamic limit, we
deduce their values by restricting us to these density regions where
the independence holds.  Then we apply the fit according to
\Eref{eq:nkall} as explained in Section \ref{sec:general}. In
\Fref{fig:corr-msr-all} we indicate by error bars the small variation
in the correlation entropy when using instead of $n_{\kappa}$ as in
\Fref{fig:nk-scaled-CF} the $n_{\kappa}$ as in the inset.  Similarly,
the corresponding variations in $N_{\rm tail}$, $S_{\rm tail}$, and
$T_{\rm tail}$ are within the symbol sizes.

\section{Extension to impenetrable Bosons and lattice gases}
\label{sec:bosons}

\begin{figure}[p]
  \centerline{\psfig{figure=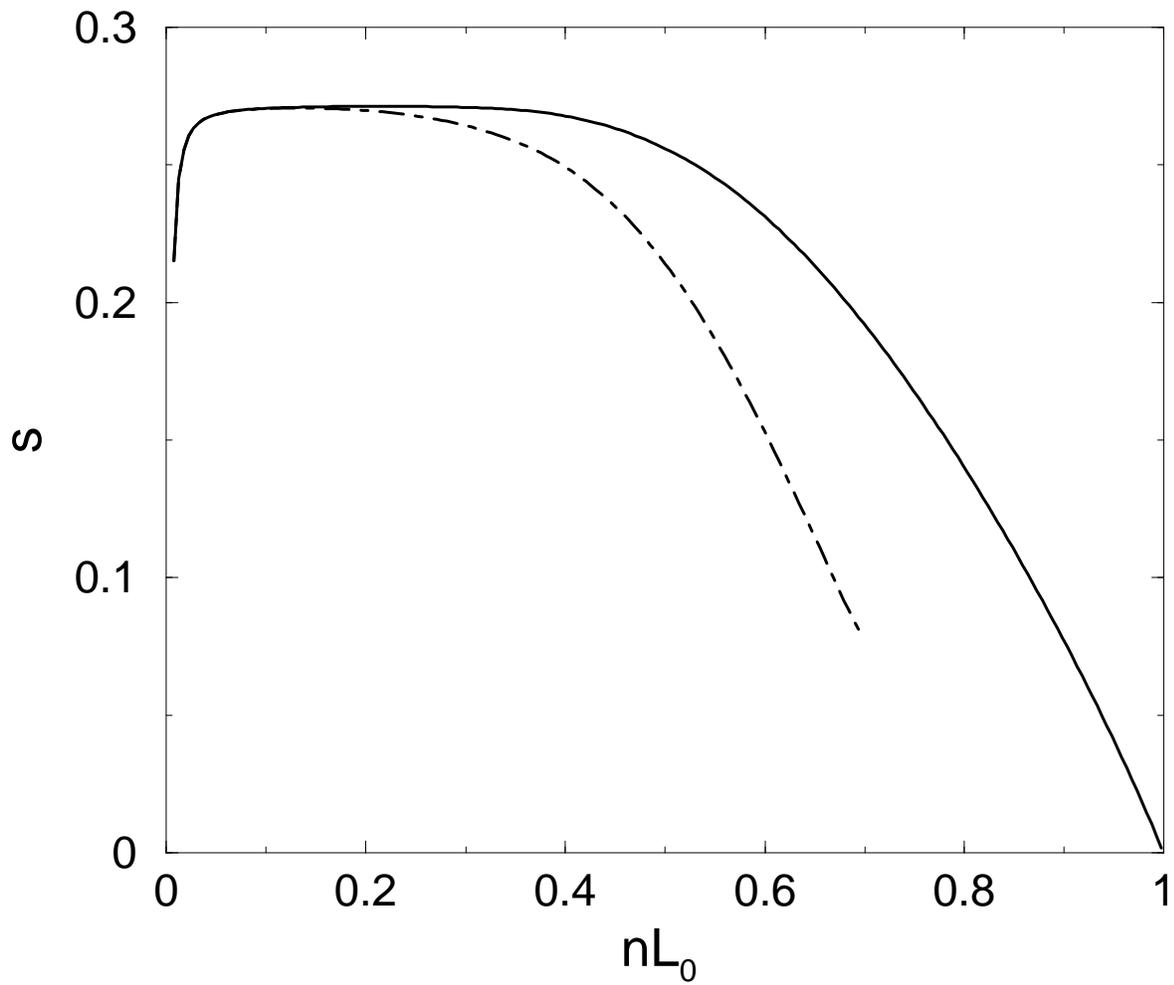,width=\columnwidth}}
  \caption{\label{fig:cb2-cf2-CE}
    Correlation entropy (\protect\ref{eq:corr-ent}) for Fermions
    (solid line) as a function of density at $\nu=3/2$. The
    dashed-dotted line corresponds to $s$ obtained for the discrete CS
    model.}
\end{figure}

\begin{figure}[p]
  \centerline{\psfig{figure=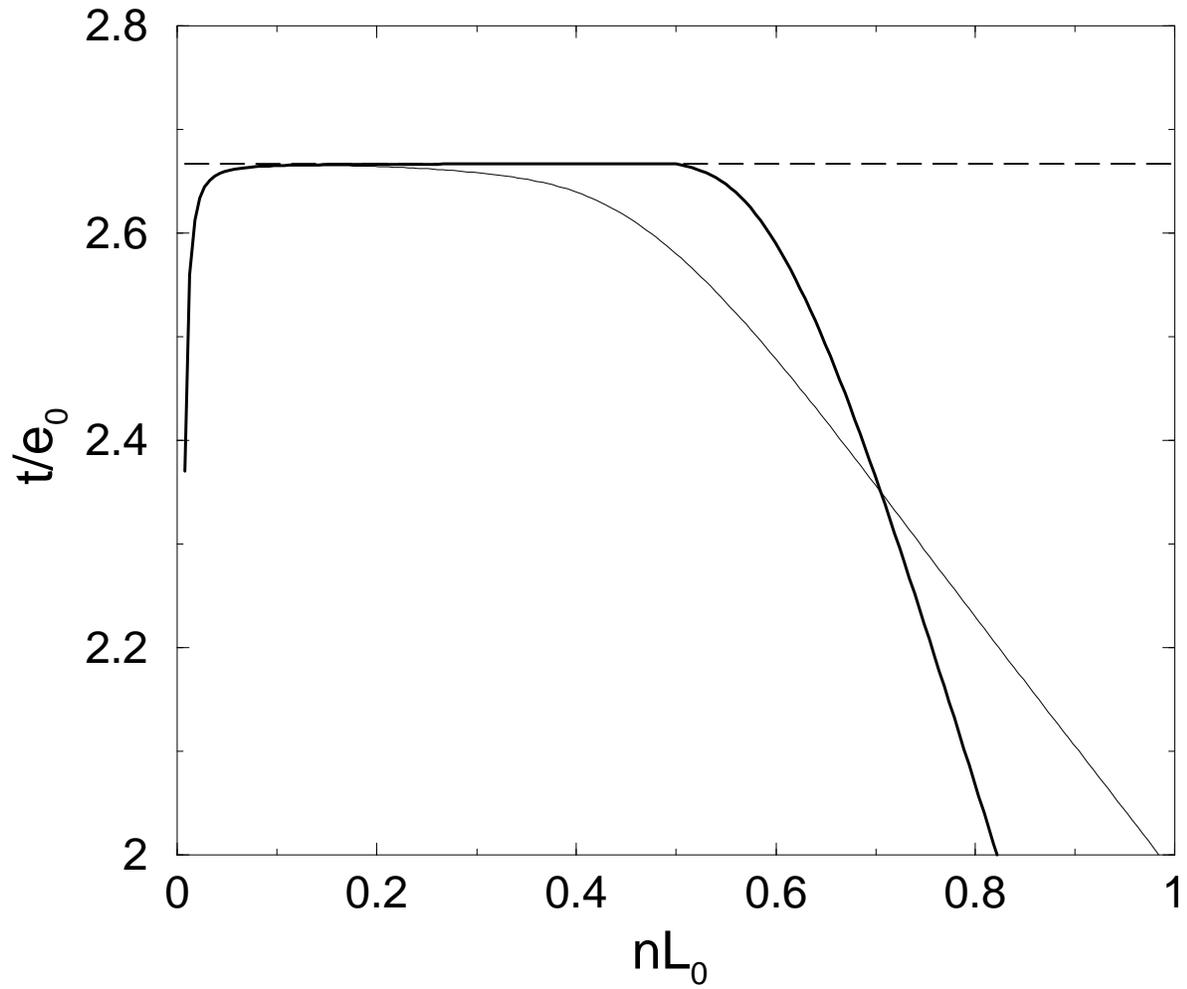,width=\columnwidth}}
  \caption{\label{fig:cb2-cf2-eng}
    Kinetic energy $t$ as computed from the Hellman-Feynman theorem
    (\protect\ref{eq:hellfeyn}) (dashed line), and $t$ from
    \Eref{eq:hellfey-r2} (solid lines) for Fermions (thick line) and
    Bosons (thin line) as a function of density at $\nu=3/2$.}
\end{figure}

\begin{figure}[p]
\centerline{\psfig{figure=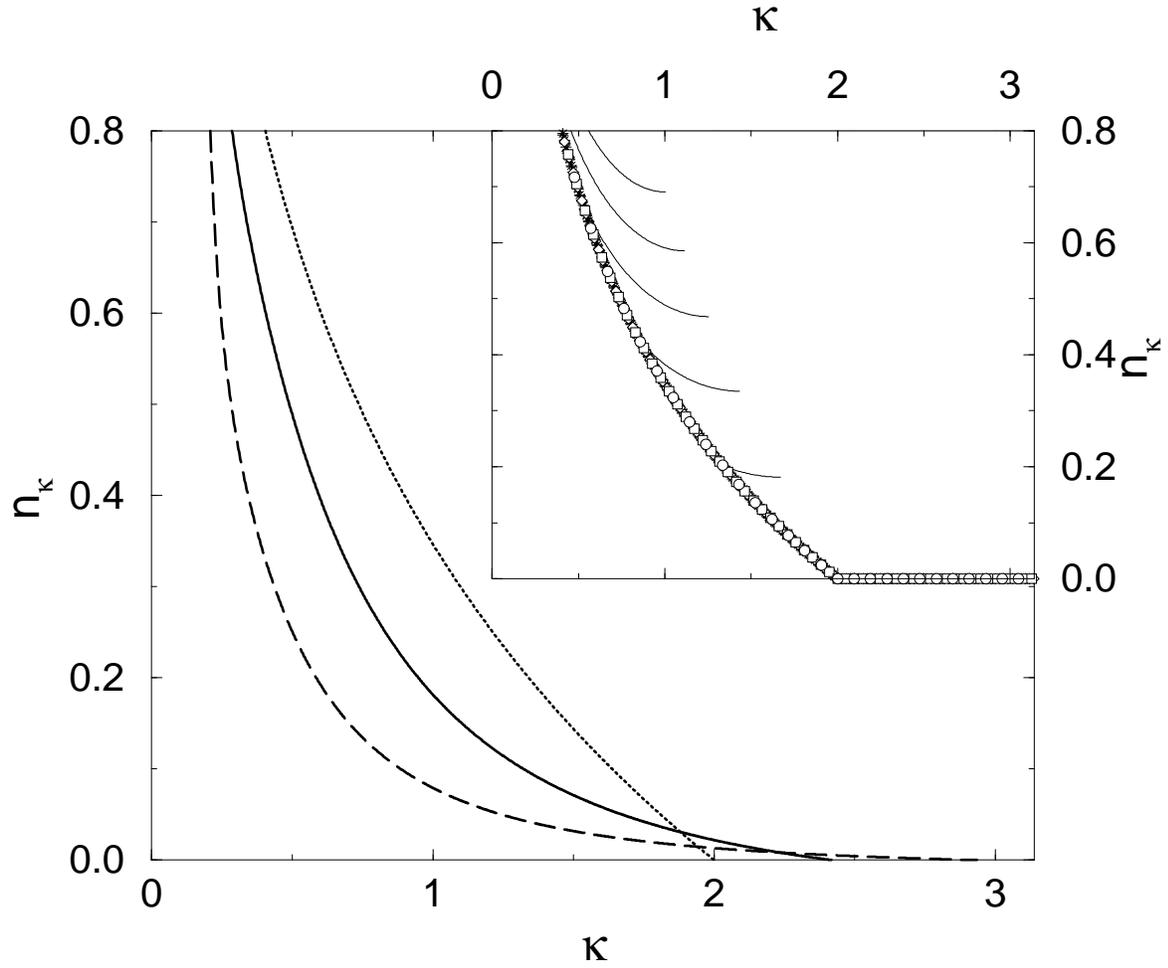,width=\columnwidth}}
   \caption{\label{fig:nk-scaled-CB}
     Bosonic momentum distributions $n_{\kappa}$ vs.\ $\kappa =
     k/k_{\rm F}$ with $\nu= 0$ (dashed), $1/2$ (solid), and $3/2$
     (dotted). Inset: $n_{\kappa}$ for $\nu= 3/2$, $L=401$ and
     particle numbers identical to the inset of
     \Fref{fig:nk-scaled-CF}.  The data for $N= 21 (\circ)$ ,
     $41(\square)$, $81(\diamond)$, $121(\times)$, $161 (+)$, and $201
     (*)$ do not show any density-dependence effects.}
\end{figure}

As mentioned in the introduction, the CS model is also solvable for
bosonic particle symmetry. The bosonic wave functions have to obey an
additional boundary condition, namely they have to vanish for
inter-particle separations $x_{ij} \to 0$ such that the resulting
system consists of {\em impenetrable} or hard-core particles
\cite{Sut71} with additional $\mu/x_{ij}^{2}$ interaction. Both PD and
1PDM may be calculated as before. The PD is independent of statistics
\cite{Sut71}, thus the fermionic exchange hole agrees with the bosonic
impenetrability hole and all quantities computed before based on the
PD are the same in the bosonic and the fermionic case.  For the 1PDM
this is different, the momentum distribution of Bosons is quite
different from the fermionic $n_\kappa$ as shown in
\Fref{fig:nk-scaled-CB}. However, energetic quantities and correlation
measures based upon those are nevertheless independent of the
statistics and should thus be the same for Bosons and Fermions. In
\Fref{fig:cb2-cf2-eng} we show that this is indeed the case. Thus
besides the density independence we have another criteria that allows
us to extract the correct values of the correlation measures from
these plots. We note that the abovementioned unwanted density
dependence is also present in the bosonic $n_\kappa$ and visible in
Figs.\ \ref{fig:cb2-cf2-CE} and \ref{fig:cb2-cf2-eng}. Also present is
the aliasing effect as shown in the inset of \Fref{fig:nk-scaled-CB}.

In Refs.\ \cite{Sut71,RomS93}, it had been shown to be useful to
restrict the family of wave functions of the CS model for both bosonic
and fermionic symmetry to a lattice such that the coordinates are
integers $x_j= 1, 2, \ldots L$ \cite{GebV87,Sha88,Hal88}. Only the
normalization constants of the wave functions change and the 1PDM can
be computed much as before \cite{Sut71}, replacing the integrals in
$F^{(\nu)}_{pq}$ by appropriate sums \cite{RomS93}. Furthermore, the
structure factor $S(q)$ is known exactly and therefore also the
PD \cite{RomS93}. The resulting lattice gas has a particle-hole
symmetry and thus we need to consider $n \leq 1/2 L_0$ only. However,
the density $N/L$ now enters all expressions in a non-trivial way and
the very useful density independence of the continuum model for the
quantities considered here is no longer applicable. Nevertheless, the
continuum model corresponds to the low-density limit of the discrete
model. In \Fref{fig:cb2-cf2-CE}, we show that this is indeed the
case for, {\em e.g.}, the correlation entropy.

\section{Discussion and Conclusions}
\label{sec:concl}

\begin{figure}[ht]
\centerline{\psfig{figure=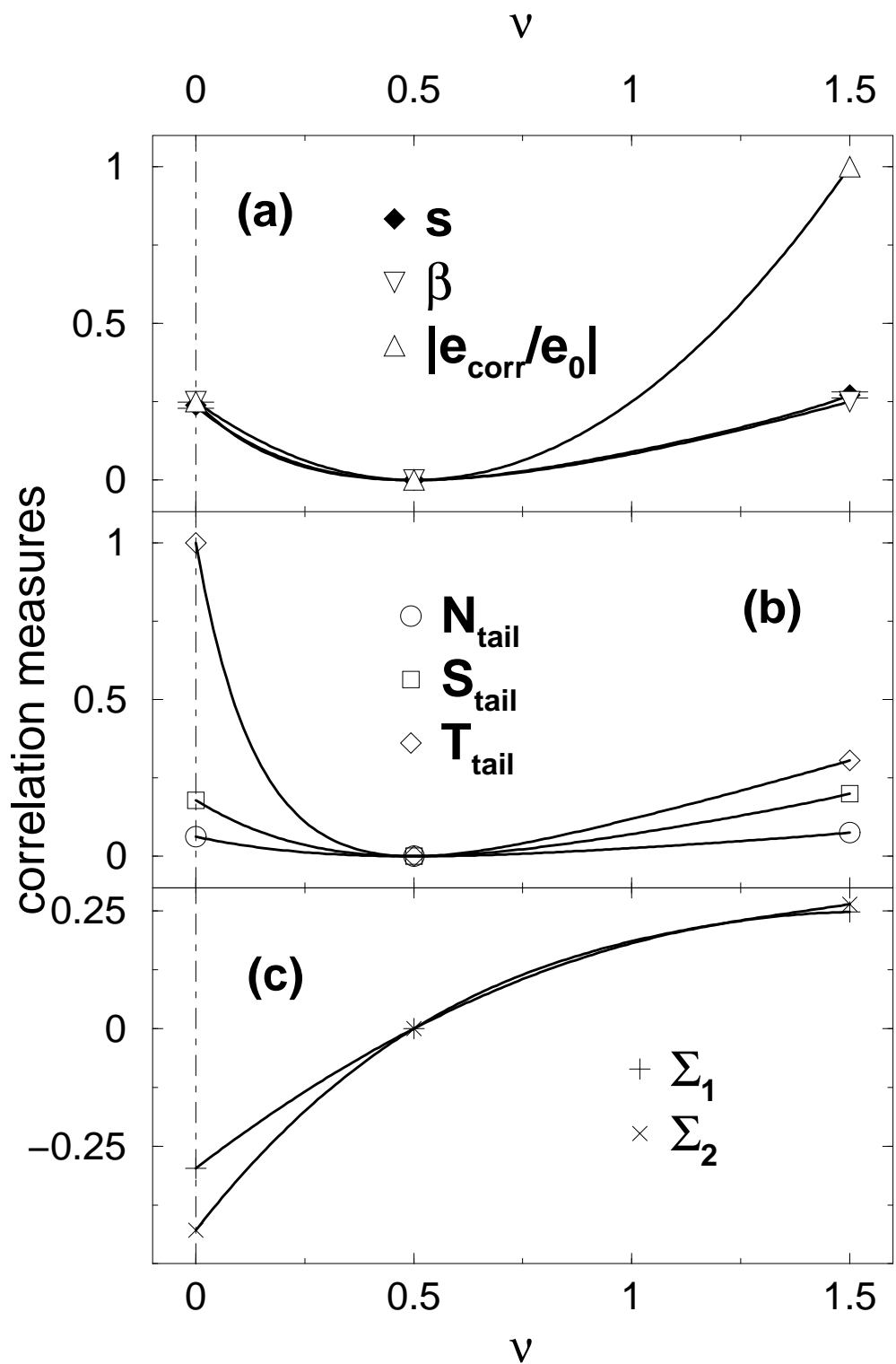,width=0.8\columnwidth}}
\caption{\label{fig:corr-msr-all}
  Correlation measures based on
  (a) the bulk of the momentum distribution, 
  (b) the tail of the momentum distribution, and
  (c) the PD
  shown as functions of $\nu$.  The solid lines are guides to the eye
  only.  The thin dashed-dotted line indicates the
  ``fall-into-the-origin'' at $\nu=0$.  }
\end{figure}

Both the PD based and the $n_\kappa$ based correlation measures
(\ref{eq:beta}) and (\ref{eq:fluc})--(\ref{eq:Ttail}) vanish for
$\nu=1/2$ (no interaction). But the first ones are more sensitive
because of $\Sigma_{1,2}\sim \nu-1/2$ near to the no-interaction point
as shown in \Fref{fig:corr-msr-all} (c), while the second ones are
$\sim (\nu-1/2)^2$ like $e_{\rm corr}$ of \Eref{fc} as shown in
Figs.\ \ref{fig:corr-msr-all} (a) and (b) and therefore
cannot distinguish between attractive and repulsive interactions. In
1D the PD based measures (\ref{eq:sig1}) and (\ref{eq:sig2}) are
identical for fermionic and (hard core) bosonic particles.  The
$n_{\kappa}$ based measures (\ref{eq:beta}) and
(\ref{eq:corr-ent})--(\ref{eq:Ttail}) do not apply for bosonic
particles, they are designed for fermionic particles only and
appropriate bosonic variants still have to be defined.

\begin{figure}[p]
  \centerline{\psfig{figure=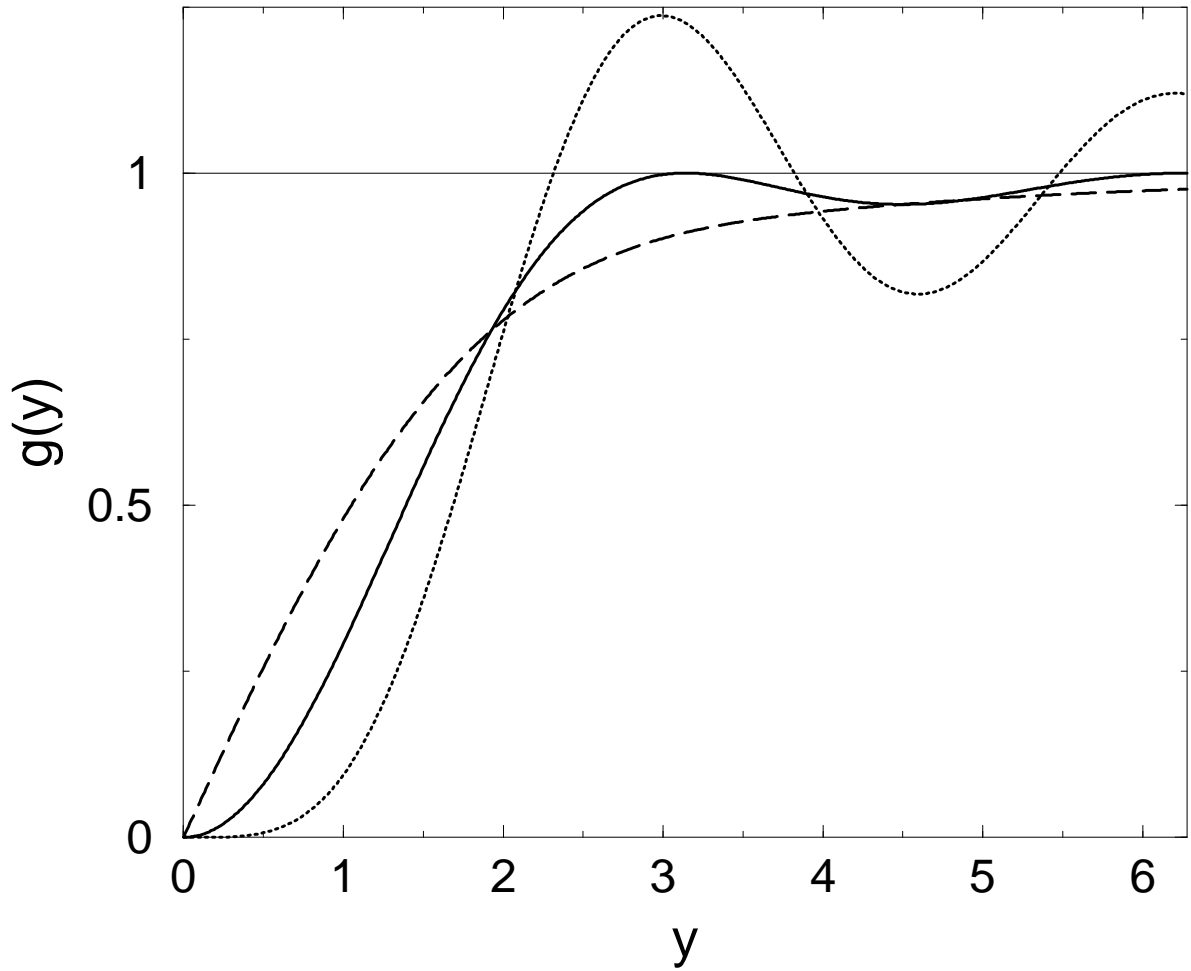,width=\columnwidth}}
   \caption{\label{fig:PD}
     Dimensionless PD $g(y)= n(x_{12})/n^2$ as a function of the
     dimensionless inter-particle separation $y= k_F x_{12}$ for $\nu=
     0$ (dashed), $1/2$ (solid), and $3/2$ (dotted). The thin line is
     a guide to the eye only.}
\end{figure}

\begin{figure}[p]
\centerline{\psfig{figure=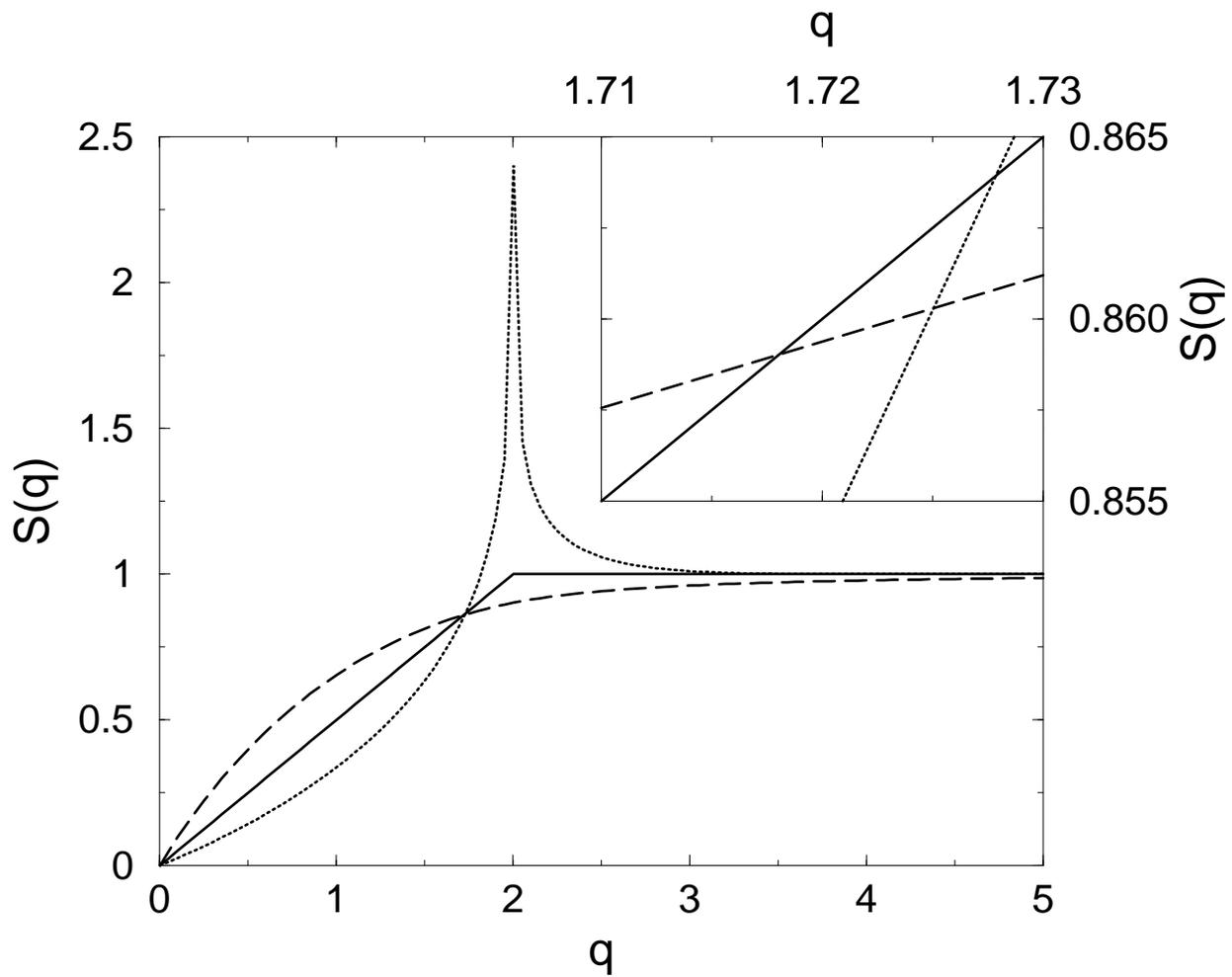,width=\columnwidth}}
\caption{\label{fig:Sq}
  Static structure factor $S(q)= 1 - \tilde{h}(q)/\pi$ for $\nu= 0$
  (dashed), $1/2$ (solid), and $3/2$ (dotted). Inset: The three curves
  do not coincide at a single point close to $q \approx 1.72$.}
\end{figure}

\begin{figure}[p]
\centerline{\psfig{figure=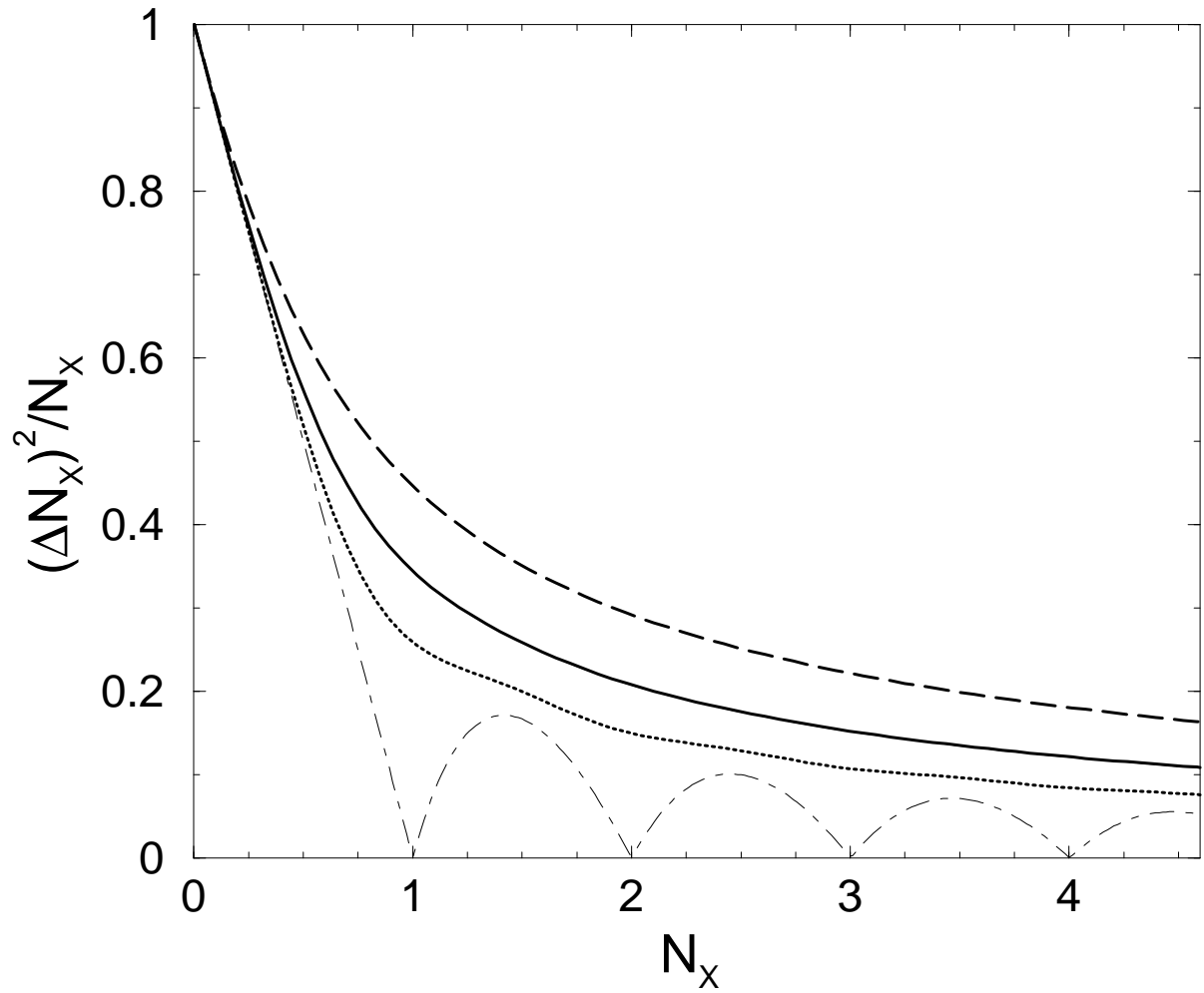,width=\columnwidth}}
\caption{\label{fig:part-num-fluc}
  Particle-number fluctuation $\left(\Delta N_X\right)^2/N_X$ in
  domains $X$ of the CS model after \Eref{eq:a42} for
  $\nu= 0$ (dashed), $1/2$ (solid), and $3/2$ (dotted). The
  dashed-dotted line corresponds to $\left(\Delta N_X\right)^2/N_X$
  for strict correlation.\protect\cite{ZieTSP99}}
\end{figure}

For repulsive particle interaction results well-known from other
extended many-body systems, such as enhancement of the Friedel
oscillations with maxima/minima trajectories (\Fref{fig:PD}),
humps/peaks of the static structure factor developing from its
non-interacting kink (\Fref{fig:Sq}), suppression of particle-number
fluctuations (\Fref{fig:part-num-fluc}), are confirmed again.
However, for attractive interactions, we have found in the present
work that particle-number fluctuations are {\em contrarily enhanced}
and that this is accompanied by a smoothening of the PD (the
oscillations disappear) and of the static structure factor (the kink
disappears) as well as by the appearance of a linear on-top behavior
of the PD. The latter behavior results in a diverging interaction
energy in the strong attraction limit although the total energy
remains finite. In momentum space the Fermi ice block thaws for both
cases and correlation tails develop. In the strong attraction limit
the correlation tail becomes so long ranged that the kinetic energy
diverges, thereby exactly compensating the divergence of the
interaction energy. We have shown that these divergences can be
derived from the exactly known energy as a function of the interaction
strength with the help of the Hellmann-Feynman theorem
(\ref{eq:hellfeyn}). This theorem allows to calculate $t(\nu)$ and
$v(\nu)$ separately from $e(\nu)$ and gives --- in addition to their
normalizations (\ref{eq:nk}) and (\ref{eq:pdd}) --- exact relations
for $n_{\kappa}$ and the PD as shown in Equations (\ref{eq:tk}) and
(\ref{eq:ie}) together with Equations (\ref{eq:gll}) and
(\ref{eq:hll}), respectively. Equations (\ref{eq:zz}) and
(\ref{eq:inr}) give additional exact integral relations between
$n_{\kappa}$ and the PD.
We expect that similar relations can be derived for correlation
function results \cite{Ha94,For95} based on the theory of Jack
symmetric polynomials for values $\nu=p/q$ with $p$ and $q$ relatively
prime positive integers \cite{Ha96}. Thus an extension of our work to
these $\nu$ is possible, albeit necessating a different approach for
the numerically stable evaluation of the products of integrals.

In summary, we have analyzed particle-number fluctuations and
studied measures for the correlation strength based on the pair
density and on the momentum distribution of the 1D quantum system of
$1/x_{ij}^2$ interacting particles.  We made extensive use of the
available exact solution and applied the Hellmann-Feynman theorem to
the CS model. Our results show that further work is called for in
order to make the qualitative terms `weak and strong correlation'
quantitatively precise presumably with a variety of quantities instead
of a single index \cite{Zie00a}.

\ack
RAR gratefully acknowledges support by the Deutsche
Forschungsgemeinschaft (SFB393) and the hospitality of the
Max-Planck-Institut f\"ur Physik komplexer Systeme (Dresden) for an
extended stay where much of this work was started. PZ thanks the
Max-Planck-Institut f\"ur Physik komplexer Systeme and P.\ Fulde for
supporting this work.



%
%

\begin{table}[h]
 \caption{\label{tab-2}
  On-top exponent and coefficients of the PD according to
  \protect\Eref{eq:kimball}.}
 \begin{indented}
\item[]\begin{tabular}{@{}l|rrr}
$\nu$ & $0$   & 1/2            & 3/2              \\
\hline
$\alpha$ & 1 & 2 & 4 \\
\hline
$A$   &$ {\pi\over 6}$ & ${1\over 3}$   & ${16\over 135}$  \\
\hline
$a_1$ & 0              & 0              & 0              \\
$a_2$ & $-{1\over 10}$ & $-{2\over 15}$ & $-{8\over 35}$   \\
$a_3$ & ${2\over {45\pi}}$ & 0          & 0 \\
$a_4$ & ${1\over {280}}$ & ${1\over {105}}$
                                        & ${32\over {1225}}$ \\
$a_5$ & $-{4\over{1575\pi}}$ & 0 & 0 \\
$a_6$ & $-{1\over {15120}}$ & $-{2\over {4725}}$
                                   & $-{2176 \over {1091475}}$ \\
$a_7$ & ${4\over {55125\pi}}$ & 0 & 0 \\
$a_8$ & ${1\over {1330560}}$ & ${2\over {155925}}$
               & ${125696}\over {1092566475}$ \\
\end{tabular}
\end{indented}
\end{table}

\begin{table}[h]
 \caption{\label{tab-1}
   Dimensionless cumulant PD $h(y)$ and the structure factor $S(q)$
   used for the computation of $\Delta N_X$ and $\Sigma_1(\nu)$ as in
   Equations  (\ref{eq:a42}) and (\ref{eq:fluc}).}
 \begin{indented}
\item[]\begin{tabular}{@{}r|l|l}
$\nu$ & $h(y)$ & $S(q)=1-{\tilde h}(q)/\pi$ \\
\hline
$0$
& $\left({\sin y\over y}\right)^2-
\left[{\rm Si}(y)-{\pi\over 2}\right]{d\over dy}{\sin y\over y}$
&$\left[q-{q\over 2}\ln (1+q)\right]\theta(2-q)$\\
& &
$+\left[2-{q\over 2}\ln {{q+1}\over {q-1}}\right]\theta (q-2)$
\\
$\frac{1}{2}$
& $\left({\sin y\over y}\right)^2$
& ${q\over 2}\theta(2-q) + \theta(q-2)$
\\
$\frac{3}{2}$
& $\left({\sin 2y\over 2y}\right)^2-{\rm Si}(2y){d\over d2y}{\sin
 2y\over 2y}$
& $\left[\frac{q}{4}-\frac{q}{8}\ln |1-{q\over 2}|\right]
\theta(4-q)$\\
 & & $+ \theta(q-4)$
\end{tabular}
\end{indented}
\end{table}

\begin{table}[h]
 \leavevmode
 \caption{\label{tab-4p5}
   Coefficients as in \Eref{eq:nk01} -- \Eref{eq:nk2i} calculated from
   the numerically determined momentum distribution $n_\kappa$ for
   $\nu=0$ and $3/2$ (at $n=1/2L_0$).}

\begin{indented}
\item[]\begin{tabular}{@{}cr|cr|cr}
\\
\multicolumn{6}{c}{$\nu= 0$}\\[1ex]
\hline
\multicolumn{2}{c}{$\kappa\in [0,1]$}
& \multicolumn{2}{c}{$\kappa\in [1,2]$}
& \multicolumn{2}{c}{$\kappa\in [2,\infty]$}\\
\hline
$B$     & 0.863355    & $B$     & 0.863355    & $C$  & 0.017788 \\
$b^-_1$ & $-$0.746775 & $b^+_1$ & $-$0.750439 &$c_2$ & 5.972791 \\
$b^-_2$ & 0.731357    & $b^+_2$ & 0.747380    &      &\\
$b^-_3$ & $-$0.420828 & $b^+_3$ & $-$0.433779 &      &\\
        &             & $b^+_4$ & 0.009552    &      &\\[3ex]
\hline
\hline\\
\multicolumn{6}{c}{$\nu= 3/2$}\\[1ex]
\hline
\multicolumn{2}{c}{$\kappa\in [0,1]$}
& \multicolumn{2}{c}{$\kappa\in [1,2]$}
& \multicolumn{2}{c}{$\kappa\in [2,\infty]$}\\
\hline
$B$     & 0.552286    & $B$     & 0.552286    & $C$  & 1.46369 \\
$b^-_1$ & 0.434380    & $b^+_1$ & 1.467126    &$c_2$ & 2.053966 \\
$b^-_2$ & $-$0.570516 & $b^+_2$ & $-$4.156361 &      &\\
        &             & $b^+_3$ & 4.180551    &      &\\
        &             & $b^+_4$ & $-$1.606130 &      &\\[3ex]
\end{tabular}
\end{indented}
\end{table}

\begin{table}[h]
\caption{\label{tab-3}
  Coefficients of the small-$X$ expansion of $\frac{\left(\Delta
    N_X\right)^2}{N_X}$ as in \Eref{eq:fluccoeff}. }
\begin{indented}
\item[]\begin{tabular}{@{}r|rrrrr}
$\nu$ & $0$                 & 1/2                   & 3/2 \\
\hline
$d_1$ & -1                  & -1                    & -1 \\
$d_2$ & ${\pi^2\over 18}$   & 0                     & 0  \\
$d_3$ & 0                   & $\pi^2\over 18$       & 0  \\
$d_4$ & $-{\pi^4\over 600}$ & 0                     & 0  \\
$d_5$ & 0                   & $-{2\pi^4\over 675}$  & ${16\pi^4\over 2025}$ \\
\end{tabular}

\end{indented}
\end{table}


%
%
\clearpage

\appendix

\section{Certain integrals}
\label{sec:integrals}

The following identities are valid with ${\rm Si}(x)=\int^x_0 dy
[\sin(y)/y]$:
\begin{eqnarray}\label{eq:a21}
\int_{0}^{\infty} \frac{dx}{\pi}
 \frac{\sin  x}{x}  = \frac{{\rm Si}(\infty)}{\pi}
& = &\frac{1}{2} \quad , \\
\label{eq:a22}
\int_{0}^{\infty} \frac{dx}{\pi}
                          { \left (
                          \frac{\sin  x}{x}
                            \right ) }^2
& = & \frac{1}{2}  \quad ,\\
\label{eq:a23}
\int_{0}^{\infty} \frac{dx}{\pi}
\left[{\rm Si}(\infty)-{\rm Si}(x)\right]
 \frac{d}{dx} \frac{\sin  x}{x}
& = & 0 \quad , \\
\label{eq:a24}
\int_{0}^{\infty} \frac{dx}{\pi}
{\rm Si}(x)
 \frac{d}{dx} \frac{\sin  x}{x}
& = & - \frac{1}{2} \quad ,\\
\label{eq:a25}
\int_{0}^{\infty} \frac{dx}{\pi}
 \frac{1}{x^2}
 \left[
  1 - { \left (
         \frac{\sin  x}{x}
        \right ) }^2
 \right]
& = & \frac{1}{3} \quad , \\
\label{eq:a26}
\int_{0}^{\infty} \frac{dx}{\pi} \frac{1}{x^2}
{\rm Si}(x)
 \frac{d}{dx} \frac{\sin  x}{x}
& = & - \frac{2}{9} \quad.
\end{eqnarray}
Equations  (\ref{eq:a22}) -- (\ref{eq:a24}) determine the normalization of
the PD's (\ref{eq:nx12}) -- (\ref{eq:nxl}). Equations  (\ref{eq:a25}) and
(\ref{eq:a26}) determine the interaction energy $v$ for the HF
approximation and for $\nu = 3/2$. For the fluctuation analysis with
Equations  (\ref{eq:partnumb}) and (\ref{eq:a42})
\begin{eqnarray}
\lefteqn{
\frac{2}{\pi}
\int_{0}^{\infty} dy \cos (qy)
\left( \frac{\sin y}{y}\right)^2 =} \nonumber \\
& & \mbox{ } \quad  \left( 1 - \frac{q}{2}\right) \theta(2 -q) \quad , \\
\lefteqn{
\frac{2}{\pi}
\int_{0}^{\infty} dy \sin (qy) \,
{\rm Si}(y)  \frac{\sin y}{y} =} \nonumber \\
& & \mbox{ } \quad  - \frac{1}{2} \ln |1-q|\  \theta(2-q) \quad , \\
\lefteqn{
\frac{2}{\pi}
\int_{0}^{\infty} dy \cos (qy) \,
{\rm Si}(y)  \frac{d}{dy} \frac{\sin y}{y} =} \nonumber \\
& & \mbox{ } \quad  - \left[ 1- \frac{q}{2} + \frac{q}{2} \ln|1-q|\right] \theta(2-q) \quad , \\
\lefteqn{
\int_{0}^{\infty} dy \cos (qy)
 \frac{d}{dy} \frac{\sin y}{y} =} \nonumber \\
& & \mbox{ } \quad  -\left[ 1- \frac{q}{2} \ln\left| \frac{1+q}{1-q} \right| \right]\quad , \\
\lefteqn{
\frac{2}{\pi}
\int_{0}^{\infty} dy \cos (qy)
\frac{\sin y}{y}  \frac{d^2}{dy^2} \frac{\sin y}{y}  =}\nonumber \\
& & \mbox{ } \quad  \frac{1}{6} (q-2)\left( q^2 -q +1 \right) \theta(2-q) \quad , \\
\lefteqn{
2\pi \int_{0}^{Y} dy_{1} \int_{0}^{Y} dy_{2}
 \left( \frac{\sin |y_1-y_2|}{|y_1-y_2|} \right)^2 =} \nonumber \\
\lefteqn{
2 \int^2_0dq\ \frac{1-\cos qY}{q^2}(1-{q\over 2})=} \nonumber \\
& & \mbox{ } \quad  1  - \cos 2Y - 2Y {\rm Si}(2Y) + \nonumber \\
& & \mbox{ } \int^{2Y}_0 dz\ \frac{1-\cos z}{z}\quad .
\end{eqnarray}

\section{Kimball like theorems for $\lowercase{n(x_{12})}$ and $\lowercase{n_\kappa}$}
\label{sec:kimball}

The small separation or on-top behavior of the PD $n(x_{12})$ is
derived here similarly as Kato \cite{Kato} and Kimball \cite{Kim75}
found the cusp relation $dg(k_{\rm F}r)/dr|_{r=0}=g(0)/a_{\rm B}$ [or
$g'(0)=\alpha r_s g(0), \alpha=(4/9\pi)^{1/3}$] for the pair
correlation of the 3D uniform electron gas.  Let us consider two
adjacent electrons with the center-of-mass and relative coordinates,
$X=(x_1+x_2)/2$ and $x=x_1-x_2$, respectively.  Focusing on the $x$
dependence the Schr\"odinger equation can be written as
\begin{equation}
\left[ -{d^2\over {dx}^2}+{\lambda(\lambda -1)\over x^2}\right]
\varphi (x){\tilde \Phi}(X,x_3,\ldots) =
(E-H')\varphi (x){\tilde \Phi}(X,x_3,\ldots) ,
\end{equation}
where $H'$ contains the remaining terms in the Hamiltonian.  Because
$E-H'$ is non-singular as $x$ approaches zero, it is unimportant for
small $x$. To lowest order in $x$ we therefore have
$\varphi(x)=x^\lambda+\ldots$, from which immediately follows
$n(x)\sim x^{2\lambda}$ for the PD, see \Eref{eq:kimball}. This can be
concluded for $\lambda \neq 1$ also directly from the many-body wave
function $\Phi\sim \Pi_{i<j} x_{ij}^\lambda$ \cite{Sut71} and for
$\lambda =1$ from \Eref{eq:g}.

A similar treatment of the asymptotic large $\kappa$ behavior of the
momentum distribution $n_\kappa$ seems to lead in \Eref{eq:nk2i}
to the conclusion $\gamma =2\lambda+2$.
This corresponds to $n_{\kappa \to \infty}\sim g(0)/\kappa ^8$ for the
3D uniform electron gas \cite{TakY91,Kim75,YasK76}.

\end{document}